\documentclass[12pt,preprint]{emulateapj}
\usepackage{graphicx}
\usepackage{rotating}

\newcommand{\br}{\ensuremath{B\!-\!R}}
\newcommand{\gr}{\ensuremath{g\!-\!r}}
\newcommand{\emax}{\ensuremath{\epsilon_{\mathrm{max}}}}

\newcommand{\amax}{\ensuremath{a_{\mathrm{max}}}}
\newcommand{\amin}{\ensuremath{a_{\mathrm{min}}}}
\newcommand{\aten}{\ensuremath{a_{10}}}
\newcommand{\lbar}{\ensuremath{L_{\mathrm{bar}}}}
\newcommand{\hi}{\ion{H}{1}}
\newcommand{\kms}{\mbox{km~s}$^{-1}$}
\newcommand{\ha}{H\ensuremath{\alpha}}

\accepted{by AJ, 5 August 2011}

\shorttitle{Outer Disks of Unbarred Galaxies}
\shortauthors{Guti\'{e}rrez et al.}

\begin{document}

\title{The Outer Disks of Early-Type Galaxies. II. Surface-Brightness Profiles of Unbarred Galaxies and Trends with Hubble Type}

\author{Leonel Guti\'{e}rrez\altaffilmark{1,2,7}, Peter Erwin\altaffilmark{3,4,5},
Rebeca Aladro\altaffilmark{2,6,7}, and John E. Beckman\altaffilmark{2,7,8}}

\altaffiltext{1}{Universidad Nacional Aut\'onoma de M\'exico, Instituto de Astronom\'ia, Ensenada, B. C. M\'exico}
\altaffiltext{2}{Instituto de Astrof\'{\i}sica de Canarias, C/ Via L\'{a}ctea s/n, 38200 La Laguna, Tenerife, Spain}
\altaffiltext{3}{Max-Planck-Insitut f\"{u}r extraterrestrische Physik, Giessenbachstrasse, 85748 Garching, Germany}
\altaffiltext{4}{Universit\"{a}ts-Sternwarte M\"{u}nchen, Scheinerstrasse 1, D-81679 M\"{u}nchen, Germany}
\altaffiltext{5}{Guest investigator of the UK Astronomy Data Centre}
\altaffiltext{6}{Present address: Department of Physics and Astronomy, University College London, Gower Street, London WC1E 6BT, UK}
\altaffiltext{7}{Facultad de F\'{\i}sica, Universidad de La Laguna, Avda. Astrof\'{\i}sico Fco. S\'{a}nchez s/n, 38200, La Laguna, Tenerife, Spain}
\altaffiltext{8}{Consejo Superior de Investigaciones Cient\'ificas, Spain}

\email{leonel@astrosen.unam.mx}

\begin{abstract} 

We present azimuthally averaged radial profiles of $R$-band surface
brightness for a complete sample of 47 early-type, unbarred galaxies, as
a complement to our previous study of early-type barred galaxies.
Following very careful sky subtraction, the profiles can typically be
determined down to brightness levels well below 27 mag arcsec$^{-2}$ and
in the best cases below 28 mag arcsec$^{-2}$. We classified the profiles
according to the scheme used previously for the barred sample: Type I
profiles are single unbroken exponential radial declines in brightness;
Type II profiles (``truncations'') have an inner shallow slope (usually
exponential) which changes at a well defined break radius to a steeper
exponential; and Type III profiles (``antitruncations'') have an inner
exponential that is steeper, giving way to a shallower outer (usually
exponential) decline.

By combining these profiles with previous studies, we can make the first
clear statements about the trends of outer-disk profile types along the
Hubble sequence (including both barred and unbarred galaxies), 
and their global frequencies.  We find that Type I
profiles are most frequent in early-type disks, decreasing from
one-third of all S0--Sa disks to barely 10\% of the latest type spirals.
Conversely, Type II profiles (truncations) \textit{increase} in
frequency with Hubble type, from only $\sim 25$\% of S0 galaxies to
$\sim 80$\% of Sd--Sm spirals. Overall, the fractions of Type I, II, and
III profiles for all disk galaxies (Hubble types S0--Sm) are: 21\%,
50\%, and 38\%, respectively; this includes galaxies ($\sim 8$\% of the total)
with composite Type II+III profiles (counted twice).

Finally, we note the presence of bars in ten galaxies
previously classified (optically) as ``unbarred''.  This suggests that $\sim 20$\%
of optically unbarred galaxies are actually barred; the bars in such
cases can be weak, obscured by dust, \textit{or} so large as to be
mistaken for the main disk of the galaxy.

\end{abstract}


\section{Introduction}\label{sec:intro}

The behavior of surface brightness --- a proxy for stellar surface
density --- in the outermost regions of galaxy disks was first discussed
in detail by \citet{vanderkruit79} and
\citet{vanderkruit81a,vanderkruit82}, who argued that disks did not
extend as a pure exponential to the limits of detection, but instead
showed a sharp or abrupt decline in surface brightness at a radius of
several scale lengths; they termed this phenomenon disk ``truncation''.
More recently, measurements in the CCD era, notably \citet{pohlen02}
using deep photometry of face-on galaxies, have shown that a more
accurate description of a truncation can be provided by a ``double'' or
``broken'' exponential fit to the radial light profile of the disk, in
which a steeper exponential breaks away from a shallower inner
exponential at a well defined galactocentric radius, the ``break
radius''.  Measurements of samples of increasing size and
representativity beyond the predominantly late-type spirals of van der
Kruit's pioneering studies have shown that there are several phenomena
in play, including extended exponential profiles without detectable
truncations \citep[e.g.,][]{weiner01,bland-hawthorn05}, as well as the
phenomenon of ``antitruncation'', where the surface-brightness profile
becomes \textit{shallower} at large radii \citep{ebp05,hunter06}.

A comprehensive classification scheme --- an extension of that
originally made by \citet{freeman70} --- has been proposed
\citep[][hereafter Paper~I]{ebp05,pt06,epb08}, in which disk profiles are grouped according
to the specific behaviour of the declining brightness in the outer disk.
This global scheme includes Type I profiles, which show a single
exponential profile extending to at least five scale lengths without
sign of truncation; Type II profiles (``truncations''), which are
truncated with a steeper outer exponential; and Type III profiles
(``antitruncations''), which are antitruncated with a shallower outer
profile. For a more detailed breakdown of this classification and some
suggested links with physical processes we refer the reader to
\citet[][hereafter PT06]{pt06} and \nocite{epb08}Paper~I.
Analysis of HST images has shown that all three types of profiles are
present in spiral galaxies out to redshifts of $z \sim 1$
\citep{perez04,trujillo05,azzollini08,bakos11}.

Theoretical interest in outer-disk profiles has traditionally focused
on the idea of explaining truncations, usually attributed to angular
momentum limits in the collapsing protogalactic cloud
\citep{vanderkruit87} or thresholds in star formation due to changes in
the gas density or phase at large radii
\citep[e.g.,][]{kennicutt89,schaye04,elmegreen06,bigiel10}. More recent studies
have examined the influence of radial diffusion driven by transient
spirals \citep{roskar08b} and cosmologically motivated simulations,
including accretion and warps \citep{foyle08,sanchez-blazquez09,martinezserrano09}.
\citet{younger07} looked at minor mergers as a possible mechanism for
creating antitruncations.

\nocite{epb08}Paper~I studied the profiles of early type
barred galaxies.  The present paper is directly complementary, in that
we confine our attention to (nominally) unbarred galaxies meeting the
same sample selection criteria. We might expect to find differences
between the two subsamples, since we understand both from Paper I and
from theoretical models by \citet{debattista06} that the presence of a
Type II profile might be related to the effects of spiral or bar
resonances; the nature and significance of such differences will be
examined in a future paper.  In any case, the
sample analyzed here taken together with the sample of Paper I forms a
representative sample of early type disks. Combined with the sample in
\nocite{pt06}PT06, which includes late type barred and unbarred spirals, it allows us to 
investigate trends with Hubble type and the global frequencies of
disk profiles.

We begin, in Section~\ref{sec:sample}, with the selection criteria for
our objects, an explanation of the sources of our images and how we have
processed them, and how we derived the surface brightness profiles. In
Section~\ref{sec:classification} we summarize the classification scheme
in sufficient detail to make the process clear to the reader.
In Section~\ref{sec:bars} we discuss
evidence for  previously undetected bars in a number of our galaxies,
and draw some conclusions about the fraction of misclassified galaxies
in standard catalogs, such as RC3. In Section~\ref{sec:discussion} we
quantify the trends in the frequencies of profile types with varying
Hubble type, the global frequencies of the different profiles, and the
similarity or lack thereof between Type I profiles and the inner and
outer components of the other profile types; in
Section~\ref{sec:conclusions} we give some brief conclusions.
Finally, Appendix~\ref{sec:profiles} presents descriptive notes on the derived
profiles, galaxy by galaxy.

\section{The galaxy sample and the data used}\label{sec:sample}

\subsection{Sample selection criteria}

The sample presented here is the unbarred counterpart to the
barred-galaxy sample presented in \nocite{epb08}Paper~I.  The parent
sample was defined as: all galaxies in the UGC \citep{nilson73} with
major axis diameter ($D_{25}$) larger than 2 arcminutes (i.e. relatively
large objects), whose ratios of major to minor axis are less than 2
(i.e. not highly inclined objects), redshift $<= 2000$ \kms{} (local),
with RC3 morphological types in the range S0 to Sb (i.e. relatively
early types), with declination above -10 degrees (i.e. observable from
the northern hemisphere). Virgo cluster galaxies, except for S0's, were
excluded to avoid uncertainties about the consistency of the Hubble
classification in Virgo \citep{vandenbergh76, koopmann98}.  The sample
here is the \textit{unbarred} subset of the parent sample: those
galaxies with RC3 classifications of SA or S (the latter formally means
that no bar classification is available), as well as those SB and SAB
galaxies which were judged to lack true bars by \citet{erwin03a} and
\citet{erwin05}. (We excluded NGC~2655 from the latter set because it
appears to be in the midst of an interaction; see
\nocite{sparke08}Sparke et al.\ 2008.) See Section~\ref{sec:bars} for more details on 
previously undetected bars.

We then removed the following galaxies because they gave strong evidence
for being edge-on systems, despite their low axis ratios: NGC~3630, 
NGC~4474, and NGC~4638. The low axis ratios generally stem from the fact
that these appear to be bulge-dominated S0 galaxies.  We also removed
NGC~4382 because we were unable to determine a clear outer-disk
orientation for this galaxy, due to strong changes in ellipticity and
position angle (without evidence for bars, rings, or spiral arms).  This
leads us to suspect that it may not be a true S0 galaxy; the fact that
it had one of the very highest fine-structure measurements in
\citet{schweizer92} suggests it may more of an elliptical-like merger
remnant.  Finally, we also removed NGC~3414, which is a peculiar system
that may be a polar-ring galaxy; the outer isophotes appear hexagonal,
and it is not clear whether the thin, bright structure bisecting the
galaxy is an extraordinarily thin bar or an edge-on disk.



In Table~\ref{T:BasicGalaxyData} we have listed the 47 galaxies in the
sample with their global properties. Most of the basic data for the
sample galaxies (coordinates, $R_{25}$, redshift, etc.) were found in
the RC3 catalog \citep{rc3}, LEDA catalogs (Lyon-Meudon Extragalactic
Database\footnote{http://leda.univ-lyon1.fr}) or from the NASA/IPAC
Extragalactic Database (NED\footnote{http://nedwww.ipac.caltech.edu}).
References for galaxy distances are given in the table.  The majority of
these are kinematic distances using the Virgocentric-inflow-corrected
velocities given by LEDA and a Hubble constant of $H_0 = 75$ \kms{}
Mpc$^{-1}$, while the next largest set of distances are
surface-brightness fluctuation measurements, mostly from
\citet{tonry01}, incorporating the suggested metallicity correction of
\citet{mei05}, which amounts to subtracting 0.06 from the distance
moduli in Tonry et al.

\begin{deluxetable*}{llrcrrrrrr}
\tablewidth{0pt}
\tablecaption{Basic Galaxy Data\label{T:BasicGalaxyData}} 
\tablecolumns{10}
\tablehead{
\colhead{Galaxy} & \colhead{Type (RC3)} & \colhead{Distance} & 
\colhead{Reference} & \colhead{$R_{25}$} & \colhead{scale} &
\colhead{PA} & \colhead{$i$} & \colhead{$M_{B}$} & \colhead{$V_{max}$} \\
 & & (Mpc) & & (\arcsec{}) & (pc/\arcsec{}) & (\arcdeg{}) & (\arcdeg) &  & (\kms{}) \\
\colhead{(1)} & \colhead{(2)} & \colhead{(3)} & \colhead{(4)} & \colhead{(5)} &
\colhead{(6)} & \colhead{(7)} & \colhead{(8)} & \colhead{(9)}  & \colhead{(10)} }

\startdata
IC~356       &    SA(s)ab pec    &   15.1  &  4  &  157  &  73   &  100   &   43  &  -21.2  &   320  \\
IC~499       &    Sa             &   28.8  &  4  &  63   &  140  &  78    &   59  &  -19.3  &   \nodata \\
NGC~278      &    SAB(rs)b       &   11.0  &  4  &  63   &  53   &  155   &   19  &  -19.4  &   181  \\
NGC~949      &    SA(rs)b        &   11.0  &  1  &  72   &  53   &  145   &   62  &  -18.5  &    95  \\
NGC~972      &    Sab            &   21.7  &  4  &  99   &  105  &  147   &   61  &  -20.4  &   151  \\
NGC~1068     &    (R)SA(rs)b     &   14.2  &  4  &  212  &  69   &  77    &   36  &  -21.2  &   198  \\
NGC~1161     &    S0$^0$         &   27.8  &  4  &  85   &  135  &  20    &   51  &  -21.1  &   \nodata \\  
NGC~2300     &    SA0$^0$        &   29.3  &  4  &  85   &  142  &  84    &   54  &  -20.6  &   \nodata \\
NGC~2460     &    SA(s)a         &   22.2  &  4  &  74   &  108  &  29    &   52  &  -19.6  &   194  \\
NGC~2775     &    SA(r)ab        &   18.0  &  4  &  128  &  87   &  165   &   39  &  -20.5  &   303  \\
NGC~2985     &    (R')SA(rs)ab   &   21.1  &  4  &  137  &  102  &  178   &   36  &  -20.7  &   237  \\
NGC~3031     &    SA(s)ab        &   3.6   &  2  &  807  &  18   &  150   &   58  &  -20.7  &   225  \\
NGC~3032     &    SAB(r)0$^0$    &   21.4  &  1  &  60   &  104  &  93    &   33  &  -18.7  &   \nodata \\ 
NGC~3169     &    SA(s)a pec     &   16.5  &  4  &  131  &  80   &  55    &   43  &  -20.2  &   391  \\
NGC~3245     &    SA(r)0$^0$     &   20.3  &  1  &  97   &  99   &  178   &   56  &  -20.0  &   \nodata \\
NGC~3455     &    (R')SAB(rs)b   &   15.8  &  4  &  74   &  77   &  62    &   57  &  -17.1  &   109  \\
NGC~3599     &    SA0$^0$        &   19.8  &  1  &  81   &  96   &  46    &   22  &  -18.7  &   \nodata \\
NGC~3604     &    SA(s)a pec     &   21.4  &  4  &  63   &  104  &  18    &   49  &  -19.3  &   173  \\
NGC~3607     &    SA(s)0$^0$     &   22.2  &  1  &  147  &  108  &  126   &   29  &  -20.9  &   \nodata \\
NGC~3619     &    (R)SA(s)0$^+$  &   23.8  &  4  &  81   &  115  &  66    &   19  &  -19.5  &   \nodata \\ 
NGC~3626     &    (R)SA(rs)0$^+$ &   19.5  &  1  &  81   &  95   &  162   &   49  &  -19.8  &   \nodata \\ 
NGC~3675     &    SA(s)b         &   12.9  &  4  &  177  &  63   &  172   &   64  &  -20.1  &   213  \\
NGC~3813     &    SA(rs)b        &   21.9  &  4  &  67   &  106  &  85    &   71  &  -19.9  &   141  \\
NGC~3898     &    SA(s)ab        &   18.9  &  4  &  131  &  92   &  107   &   53  &  -20.5  &   274  \\
NGC~3900     &    SA(r)0$^+$     &   25.9  &  4  &  95   &  126  &  1     &   61  &  -20.0  &   \nodata \\ 
NGC~3998     &    SA(r)0$^0$?    &   13.7  &  1  &  81   &  67   &  136   &   38  &  -19.4  &   \nodata \\ 
NGC~4138     &    SA(r)0$^+$     &   13.4  &  1  &  77   &  65   &  150   &   55  &  -18.4  &   \nodata \\ 
NGC~4150     &    SA(r)0$^0$     &   13.7  &  1  &  70   &  66   &  146   &   50  &  -18.3  &   \nodata \\
NGC~4223     &    SA(s)0$^+$     &   16.5  &  2  &  61   &  80   &  126   &   59  &  -18.2  &   \nodata \\ 
NGC~4281     &    S0$^+$         &   23.8  &  1  &  91   &  115  &  86    &   66  &  -19.7  &   \nodata \\
NGC~4369     &    (R)SA(rs)a     &   16.6  &  4  &  63   &  81   &  53    &   19  &  -18.8  &   134  \\
NGC~4459     &    SA(r)0$^+$     &   16.1  &  3  &  106  &  78   &  102   &   38  &  -19.8  &   \nodata \\
NGC~4578     &    SA(r)0$^0$     &   16.3  &  3  &  99   &  79   &  31    &   47  &  -18.8  &   \nodata \\
NGC~4736     &    (R)SA(r)ab     &   5.1   &  1  &  337  &  25   &  119   &   30  &  -20.0  &   194  \\
NGC~4750     &    (R)SA(rs)ab    &   25.4  &  4  &  61   &  123  &  173   &   42  &  -20.3  &   193  \\
NGC~4772     &    SA(s)a         &   14.5  &  4  &  102  &  70   &  147   &   46  &  -19.2  &   279  \\
NGC~4826     &    (R)SA(rs)ab    &   7.3   &  1  &  300  &  35   &  113   &   61  &  -20.6  &   153  \\
NGC~4880     &    SA(r)0$^+$     &   19.7  &  4  &  95   &  95   &  159   &   43  &  -18.5  &   \nodata \\
NGC~4941     &    (R)SAB(r)ab    &   15.0  &  4  &  109  &  73   &  21    &   48  &  -19.3  &   185  \\
NGC~5273     &    SA(s)0$^0$     &   16.1  &  1  &  83   &  78   &  9     &   31  &  -18.6  &   \nodata \\
NGC~5485     &    SA0$^0$ pec    &   25.2  &  1  &  70   &  122  &  167   &   49  &  -19.7  &   \nodata \\ 
NGC~5520     &    Sb             &   28.5  &  4  &  60   &  138  &  64    &   62  &  -19.5  &   138  \\
NGC~6340     &    SA(s)0/a       &   20.2  &  4  &  97   &  98   &  175   &   20  &  -19.8  &   \nodata \\ 
NGC~7217     &    (R)SA(r)ab     &   14.9  &  4  &  117  &  72   &  89    &   28  &  -20.3  &   306  \\
NGC~7457     &    SA(rs)0$^-$?   &   12.9  &  1  &  128  &  62   &  126   &   58  &  -18.9  &   \nodata \\
UGC~3580     &    SA(s)a pec     &   19.3  &  4  &  102  &  94   &  7     &   64  &  -18.7  &   104  \\
UGC~4599     &    (R)SA0$^0$     &   28.2  &  4  &  60   &  137  &  91    &   24  &  -17.4  &   \nodata \\ 
\enddata

\tablecomments{Basic data for the galaxies in our sample. Col.\ (1) Galaxy
name; (2) Classification from \citet{rc3} (RC3); (3) Distance in Mpc; (4)
Reference for distance: 1 = \citet{tonry01}, including metallicity correction
from \citet{mei05}, 2 = \citet{freedman01}, 3 = mean distance to Virgo Cluster
from \citet{mei07}, 4 = radial velocity, corrected for Local Group infall onto 
Virgo (from LEDA), using $H_0 = 75$ km s$^{-1}$ Mpc$^{-1}$; (5) Half of the 
corrected $\mu_{B} = 25$ magnitude diameter $D_{0}$ from RC3; (6) Scale in pc
arcsec$^{-1}$; (7) and (8) Position angle and inclination of the outer disk,
measured in this work (see Section~\ref{sec:ellipse_fitting}); 
(9) Absolute $B$ magnitude, using the corrected
apparent magnitude $B_{\rm tc}$ from LEDA and the distance in column 3; 
(10) Maximum rotation velocity calculated using the apparent maximum rotation 
velocity of gas $V_{maxg}$ from LEDA corrected for inclination in column 8.}

  
\end{deluxetable*}

We note that after a detailed analysis of the images, a number of the
``unbarred'' galaxies turned out to have bars (and a fraction of these even
have two bars!); Section~\ref{sec:bars} discusses the individual cases and
Table~\ref{T:bars} presents the bar parameters for these galaxies. 
Despite the fact that these galaxies have proved to be barred rather than 
unbarred, we included them in our full outer-disk analysis (though they are 
grouped with other barred galaxies when we discuss population properties).

\subsection{Sources of the images used}

In Table~\ref{T:Observations}, we give a list of the galaxies and relevant information about the observations and calibrations.  This table lists the image sources used for generating the surface-brightness profiles, along with the exposure time and calibration method used.

\begin{deluxetable*}{lllcll}
\tablewidth{0pt}
\tablecaption{Observations and Calibrations\label{T:Observations}} 
\tablecolumns{6}
\tablehead{
\colhead{Galaxy} & \colhead{Telescope/Instrument} & \colhead{Date}
& \colhead{$t_{\rm exp}$ (s)} & \colhead{Calibration} &
\colhead{Notes} }
\startdata
IC~356   &  INT-WFC     &    2004-03-16  &      2$\times$600   &  standards &  \\ 
IC~499   &  INT-WFC     &    2004-03-16  &      2$\times$600   &  standards &  \\
NGC~278 &  INT-WFC      &    2004-12-11  &      3$\times$ 60   &  standards & 1, 2 \\
NGC~949 &  INT-PFCU     &    1994-12-04  &      240   &     PH98 & 2    \\
NGC~972 &  JKT/JAG      &    2000-11-01  &      2$\times$1200    &  J04     & 2  \\ 
NGC~1068 &  SDSS        &       DR5      &      54    &  SDSS &     \\
NGC~1161 &  INT-WFC     &    2004-03-16  &      30    &  standards & \\
NGC~2300 &  INT-WFC     &    2004-03-16  &      60    &  standards & \\
NGC~2460 &  INT-WFC     &    2004-03-14  &     2$\times$600    &  standards & \\
NGC~2775 &  SDSS        &      DR5       &      54    &  SDSS &     \\
NGC~2985 &  JKT/JAG     &    2003-03-08  &     3$\times$300    &  N07   &    \\ 
NGC~3031 &  SDSS        &      DR5       &      54    &  SDSS  &    \\
NGC~3032 &  SDSS        &      DR5       &      54    &  SDSS  &    \\
NGC~3169 &  INT-WFC     &    2004-03-14  &     2$\times$600    &  standards & \\
NGC~3245 &  INT-WFC     &    2004-03-16  &      60    &  standards & \\
NGC~3455 &  SDSS        &      DR5       &      54    &  SDSS &     \\
NGC~3599 &  INT-WFC     &    2004-03-14  &     2$\times$600    &  standards & \\
NGC~3604 &  INT-WFC     &    2004-03-15  &     2$\times$600    &  standards & \\
NGC~3607 &  SDSS        &      DR5       &      54    &  SDSS &     \\
NGC~3619 &  SDSS        &      DR5       &      54    &  SDSS &     \\
NGC~3626 &  SDSS        &      DR5       &      54    &  SDSS &     \\
NGC~3675 &  SDSS        &      DR5       &      54    &  SDSS &     \\
NGC~3813 &  SDSS        &      DR5       &      54    &  SDSS &     \\
NGC~3898 &  SDSS        &      DR5       &      54    &  SDSS &     \\
NGC~3900 &  INT-WFC     &    2004-03-14  &     600    &  standards & \\
NGC~3998 &  INT-WFC     &    2004-03-16  &      60    &  standards & \\
NGC~4138 &  SDSS        &      DR5       &      54    &  SDSS &     \\
NGC~4150 &  INT-WFC     &    2004-03-17  &      60    &  standards & \\
NGC~4223 &  SDSS        &      DR5       &      54    &  SDSS &     \\
NGC~4281 &  SDSS        &      DR5       &      54    &  SDSS &     \\
NGC~4369 &  INT-WFC     &    2004-03-15  &     2$\times$600    &  standards & \\
NGC~4459 &  SDSS        &      DR5       &      54    &  SDSS &     \\
NGC~4578 &  SDSS        &      DR5       &      54    &  SDSS &     \\
NGC~4736 &  INT-WFC     &    2004-03-17  &     120    &  standards & \\
NGC~4750 &  INT-WFC     &    2004-03-17  &     2$\times$600    &  standards & \\
NGC~4772 &  SDSS        &      DR5       &      54    &  SDSS &     \\
NGC~4826 &  INT-WFC     &    2004-03-17  &     2$\times$600    &  standards & \\
NGC~4880 &  SDSS        &      DR5       &      54    &  SDSS &     \\
NGC~4941 &  INT-WFC     &    2004-03-15  &     2$\times$600    &  standards & \\
NGC~5273 &  INT-WFC     &    2004-03-16  &     2$\times$600    &  standards & \\
NGC~5485 &  INT-WFC     &    2004-03-15  &     2$\times$600    &  standards & \\
NGC~5520 &  SDSS        &      DR5       &      54    &  SDSS &     \\
NGC~6340 &  INT-WFC     &    2004-03-14  &     600    &  standards & \\
NGC~7217 &  JKT/JAG     &    2000-05-31  &     4$\times$600    &  K06       & \\ 
NGC~7457 &  JKT/JAG     &    2002-09-09  &     3$\times$600    &    PH98  & 2, 3   \\
UGC~3580 &  INT-WFC     &    2004-03-14  &     2$\times$600    &  standards &  \\
UGC~4599 &  INT-WFC     &    2004-03-17  &     2$\times$1200    &  standards & 4 \\
\enddata
\tablecomments{In the Calibration column, ``standards'' indicates calibration using simultaneous observations of Landolt standard stars, while PH98 indicates use of aperture photometry from \citet{prugniel98}; J04, calibrations from \citet{james04}; N07, calibrations from \citet{noordermeer07}; and K06, calibrations from \citet{kassin06}.  Notes: 1 = Calibration based on \citet{knapen-n278}; 2 = archival data; 3 = $V$-band image calibrated to Cousins $R$ (see text); 4 = $B$-band image only.}
\end{deluxetable*}

The largest subset of images (22 galaxies) 
came from an observing run with the Wide Field Camera (WFC) of the Isaac Newton Telescope (INT) of the Roque de los Muchachos Observatory (ORM) La Palma, Spain, in the period 14th--17th March 2004. The conditions were photometric, although the seeing varied considerably (between 0.8\arcsec{} and 3.4\arcsec{}).  Even the worst seeing in this run does not, however, pose a real problem for our work, since we treat data from the outer parts of the disks where averages over relatively large areas are used.  All of the INT-WFC images used in this work were acquired with the $r$ filter, with one exception: UGC 4599.  We did make observations of this galaxy in the $r$ band in a previous run (December 2003); however, the resulting images were dominated by scattered light and not nearly as useful as the $B$-band image acquired in March 2004.  Consequently, we used the latter for this galaxy.\footnote{Analysis of the $r$-band images does result in a similar profile, albeit more limited by sky-background uncertainty.}

For 19 of the galaxies, 
we used Data Release 5 (DR5) of the Sloan Digital Sky Survey (SDSS\footnote{http://www.sdss.org/dr5/}) \citep{york00, adelman07}.  As discussed in \nocite{pt06}PT06 and \nocite{epb08}Paper~I, the highly uniform sky background of the SDSS images means that we can use them to derive profiles extending further out than might naively be thought, given the relatively short exposure times.

Images for six galaxies were taken from the archive of the Isaac Newton Group at the Astronomical Data Centre of the Cambridge Astronomical Search Unit (CASU)\footnote{http://casu.ast.cam.ac.uk/casuadc/archives/ingarch}. Four of these were taken with the Jacobus Kapteyn Telescope (JKT) and two with the Isaac Newton Telescope.  We used images taken with $R$ or $r$ filters for all but one of these galaxies; the exception was NGC~7457, for which the best available image was in the $V$ band calibrated to $R$ (see Section~\ref{sec:calibration}).

For those objects where we have both SDSS and INT-WFC images we have used the latter, as they are deeper, except for those few cases where there are strong background variations or other imperfections. 

\subsection{Image processing}

\subsubsection{Image preparation}

The images we took from the SDSS were already reduced. However, in some cases we had to merge adjacent fields from the same imaging run in order to obtain the complete image of the galaxy. Only for NGC~3032 did we merge images from different runs, photometrically calibrating each image separately, subtracting the sky background, and correcting for the factor equivalent to the zero point difference before merging (see Section~\ref{sec:calibration} for more details on the calibration). 

For the INT-WFC images we had to reduce the raw images; we flat-fielded and bias-corrected them, and corrected for the non-linearity of the WFC
CCDs found by the Isaac Newton Group and the Cambridge Astronomical Survey Unit's INT Wide Field Survey.\footnote{See http://www.ast.cam.ac.uk/~wfcsur/technical/foibles/.}
Alignment of multiple exposures was performed using standard IRAF\footnote{Image Reduction and Analysis Facility (IRAF); http://iraf.noao.edu} tasks.
After alignment and sky background correction the images were combined using the IRAF task \emph{imcombine}, which was done for all cases except those few for which only a single exposure was available. For all but two of the INT-WFC observations, the galaxies fit inside the central CCD chip (Chip 4 of the mosaic), and so we did not need to create a mosaic.  The exceptions were NGC~4736 and NGC~4826, which were large enough to overfill Chip 4. For these galaxies, we created mosaic images by copying the individual chip images into a single image, using appropriate rotations and offsets as determined by \citet{eliche-moral06}.

Archival images taken with the JKT and the INT-PFCU were reduced in similar fashion.

\subsubsection{Sky subtraction and image masking}\label{sec:sky_subtraction}

A key part of the reduction of our images was the sky subtraction, which is of critical importance when measuring the faint outer parts of the disk. We used the technique described in detail in \nocite{epb08}Paper~I. The SDSS images were easier to handle because their sky backgrounds were highly uniform. However, for all the galaxies in the sample we first probed the background for large scale variations, applying a median filter with a 9$\times$9 pixel box across the whole image. If a gradient was present, we used the IRAF task \emph{imsurfit} to correct the original image. For using this task we selected rectangular regions of the sky that did not include galaxies or bright stars.
 
Following this, we subtracted off a background level whose value was measured in clean areas of the image far from the edge of the detectable surface brightness increment due to the galaxy, and avoiding cosmetic chip defects, stars and field galaxies. A typical background estimate used median values from a set of 70--100 boxes, each 10$\times$10 pixels in size.  The mean value of the individual medians was then calculated; the uncertainty on this mean $\sigma_{\rm sky}$ was calculated by bootstrap resampling.  (See \nocite{epb08}Paper~I for a more detailed discussion of this methodology.)  As was done by \nocite{pt06}PT06 and \nocite{epb08}Paper~I, we derive a confidence limit for our surface brightness profiles corresponding to 4.94 $\sigma_{\rm sky}$, which is the surface brightness level where an error of one $\sigma_{\rm sky}$ in the background measurement would change the profile by 0.2 magnitudes arcsec$^{-2}$. In our graphical presentations of the profiles (Figure~\ref{fig:profiles}; see Section~\ref{sec:ellipse_fitting}) we have plotted them until they become obviously noisy, in general down to a little below this defined uncertainty limit, which typically lies between 26.0 and 28.0 mag arcsec$^{-2}$. We note that 8 
of the galaxies in the sample have uncertainty limits at values fainter than 28.0 mag arcsec$^{-2}$.

The masking process is an important feature of our procedure. It entails flagging those regions which contain bright stars, or field galaxies, or instrumental defects (reflections, ghosts, bad columns etc.). The flagged regions, marked out as circles, ellipses, rectangles or polygons via the DS9 image display, are converted to IRAF pixel-list (.pl) format mask images, which the ellipse-fitting routine uses to identify masked pixels. We found that in general the INT-WFC images required more masking than those from the SDSS, and galaxies closer to the Galactic plane also required considerable masking.

\subsubsection{Image Calibration}\label{sec:calibration} 

To maintain as uniform a calibration scheme as possible we have standardized our images on the Cousins $R$ band. For many galaxies we used standard stars as calibrators, in others we calibrated indirectly using aperture photometry from the literature, and for the SDSS images we converted the SDSS $r$ zero points to Cousins $R$. 
To do that, as in \nocite{epb08}Paper~I, we determined the appropriate Cousins zero point via the following expression from \citet{smith02}:

\begin{equation}
\mathrm{ZP}_R  \, = \, \mathrm{ZP}_r - 0.14 (\gr) - 0.14
\end{equation}

\noindent where ZP$_R$ is the Cousins \emph{R} zero point and \gr{} is the corresponding color for the galaxy (see \nocite{epb08}Paper~I for details).  
We determined \gr{} with circular apertures, using the largest apertures possible which did not extend beyond the main body of the galaxy. The photometry used  the IRAF task \emph{apphot} from the \emph{digiphot} package. Table~\ref{T:CousinsRforSDSS} lists the \gr{} colors and the resulting Cousins $R$ zero points for the set of galaxies taken from the SDSS.

\begin{deluxetable}{lllrrll}
\tablewidth{0pt}
\tablecaption{Cousins R Calibrations for SDSS Images\label{T:CousinsRforSDSS}} 
\tablecolumns{5}
\tablehead{
\colhead{Galaxy} & \colhead{ZP$_r$} & \colhead{ZP$_g$} & 
\colhead{\gr} & \colhead{ZP$_R$} 
}
\startdata
NGC~1068  &  26.25  &   26.55  &  0.64  &   26.02    \\
NGC~2775  &  26.25  &   26.61  &  0.77  &   26.01    \\
NGC~3031  &  26.20  &   26.56  &  0.81  &   25.95    \\
NGC~3032  &  26.19  &   26.54  &  0.54  &   25.97    \\
NGC~3455  &  26.15  &   26.44  &  0.39  &   25.95    \\
NGC~3607  &  26.14  &   26.43  &  0.76  &   25.89    \\
NGC~3619  &  26.23  &   26.53  &  0.73  &   25.99    \\
NGC~3626  &  26.25  &   26.69  &  0.67  &   26.02    \\
NGC~3675  &  26.22  &   26.53  &  0.79  &   25.97    \\
NGC~3813  &  26.23  &   26.52  &  0.51  &   26.02    \\
NGC~3898  &  26.23  &   26.67  &  0.75  &   25.99    \\
NGC~4138  &  26.19  &   26.56  &  0.74  &   25.95    \\
NGC~4223  &  26.22  &   26.52  &  0.76  &   25.97    \\
NGC~4281  &  26.25  &   26.57  &  0.79  &   26.00    \\
NGC~4459  &  26.27  &   26.73  &  0.83  &   26.01    \\
NGC~4578  &  26.26  &   26.58  &  0.77  &   26.02    \\
NGC~4772  &  26.23  &   26.48  &  0.71  &   25.99    \\
NGC~4880  &  26.20  &   26.58  &  0.66  &   25.97    \\
NGC~5520  &  26.22  &   26.67  &  0.52  &   26.01    \\
\enddata
\tablecomments{Cousins $R$ calibrations for galaxies with images taken from the SDSS. Columns: (1) Galaxy name; (2) and (3) SDSS zero points calculated using the associated \emph{tsField} tables; (4) \gr{} color determined from aperture photometry; (5) zero points for Cousins-$R$ magnitude, found as described in Section \ref{sec:calibration}}
\end{deluxetable}

The $R$-band calibrations for our 2004 INT-WFC run were based on simultaneous standard star 
observations and were presented in \nocite{epb08}Paper~I. These calibrations require an observed \br{} 
color for each galaxy. In most cases, we used \br{} colors from the aperture photometry
collected in \citet{prugniel98}; for NGC~5273 we used the value published by \citet{barway05}. 
For those galaxies where we could not find a published value, we assumed the following mean (unreddened) values 
for the different Hubble types (see \nocite{epb08}Paper~I): S0: \br{} = 1.5; S0/a \& Sa: \br{} 
= 1.4; Sab \& Sb: \br{} = 1.3. We then applied Galactic reddening based on the values in NED \citep[taken from][]{schlegel98} to 
estimate what the observed \br{} color would be for those galaxies. In Table~\ref{T:ColorsWFC} we list 
the galaxies observed with the INT-WFC and their corresponding \br{} values.

\begin{deluxetable}{llllll}
\tablewidth{0pt}
\tablecaption{Colors for Galaxies Observed with INT-WFC\label{T:ColorsWFC}} 
\tablecolumns{6}
\tablehead{
\colhead{Galaxy} & \colhead{\br} & \colhead{Galaxy} & 
\colhead{\br} & \colhead{Galaxy} & \colhead{\br} }
\startdata
IC~356   & 2.27 &  NGC~3604 & 1.48 &  NGC~4826 & 1.37 \\
IC~499   & 1.51 &  NGC~3900 & 1.55 &  NGC~4941 & 1.41 \\
NGC~1161 & 1.86 &  NGC~3998 & 1.53 &  NGC~5273 & 1.31 \\
NGC~2300 & 1.67 &  NGC~4150 & 1.53 &  NGC~5485 & 1.62 \\
NGC~2460 & 1.52 &  NGC~4369 & 1.44 &  NGC~6340 & 1.53 \\
NGC~3169 & 1.46 &  NGC~4736 & 1.33 &  UGC~3580 & 1.49 \\
NGC~3245 & 1.53 &  NGC~4750 & 1.38 &  UGC~4599 & 1.55 \\
NGC~3599 & 1.25 &  \\ 
\enddata              
\end{deluxetable}

Images of the galaxies NGC~278, NGC~949, NGC~972, NGC~2985, NGC~7217, and NGC~7457 were obtained from the ING archive and did not, in general, have accompanying standard-star observations.  We calibrated these images using aperture photometry (with the IRAF task \emph{apphot}), compared with published aperture data: For NGC~278 we used photometric data from \citet{knapen-n278}, for NGC~949 we used data from \citet{heraudeau96}, for NGC~972 two aperture measurements kindly provided by Phil James (private communication, based on data from \nocite{james04}James et al.\ 2004), for NGC~2985 the ``$R(m_{25})$''\footnote{The $R$ magnitude integrated out to $D_{25}$, the isophotal diameter at 25 mag arcsec$^{-2}$.} value of \citet{noordermeer07}, and for NGC~7217 the Cousins-\emph{R} magnitude of \citet{kassin06}.
 
In the case of NGC~7457, the best available image was actually a $V$-band observation from the ING Archive.  We transformed this to the $R$-band using aperture photometry from \citet{prugniel98}.  Since the \bv{} and \br{} color profiles for this galaxy in \citet{michard00} are almost flat, this is probably a reasonable approach, and the true $R$-band profile will not be significantly different from ours.

\subsection{Ellipse fitting and deriving the surface brightness profiles}\label{sec:ellipse_fitting}
 
Radial surface brightness profiles can be constructed by fitting ellipses to the galaxy isophotes. There are two basic ways to do this.  The first (``free'' fitting) leaves the ellipticity and position angle of the ellipses as free variables, along with the mean surface brightness, to be determined for each value of the semi-major axis. The second method (``fixed'' fitting) fixes the values of the ellipticity and position angle to that of the outer disk, so that only the mean surface brightness is a free parameter.  The latter method has the virtue that one is effectively averaging around concentric circles of the deprojected galaxy. Distortions due to smaller scale features (e.g. spiral arms) tend to be smoothed out, although if prominent they can still be traced in the global profile. This technique was used effectively by \nocite{pt06}PT06 and in \nocite{epb08}Paper~I; see the latter for discussion of why free-ellipse fitting can produce distortions in surface-brightness profiles if strong non-axisymmetric structures are present. 

Our basic approach is the same as in \nocite{epb08}Paper~I: we use
free-ellipse fitting to help determine the orientation (apparent
ellipticity and position angle) of the outer disk. The values given in
Table~\ref{T:BasicGalaxyData} are the results of this
process. That table lists inclination $i$ for each galaxy, derived
from the outer-disk ellipticity $\epsilon$ using the formula \citep{hubble26}

\begin{equation}
cos^2i = \frac{(1 - \epsilon)^2 - q_o^2}{1 - q_o^2},
\end{equation}

\noindent where $q_o$ is the intrinsic flattening of the disk, for which we assumed 
that the outer disk is an
axisymmetric ellipsoid with axis ratio $q_o = c/a = 0.2$ \citep{degrijs98}. Once this is done,
we re-run the ellipse-fitting software (the \textit{ellipse} task in
IRAF) in ``fixed'' mode, with ellipticity and position angle held
constant to the values of the outer disk. When deriving the surface
brightness profiles using fixed fitting we took a logarithmic radial
sampling interval, with steps which increase the radius by 3\% each. The
surface brightness value obtained is the median value for the pixels in
the given elliptical annulus. Using the median helps ensure that cosmic
rays, bad pixels, faint stars, etc., are not included in the result.

We fitted exponentials to those parts of the radial surface brightness profiles where the plots, in magnitude \emph{vs} linear radius, are clearly linear. Tests using a subsample 
of galaxies suggest that typical uncertainties due to the exact choice of fitting range are  $\sim 0.1$ mag for the central surface brightnesses $\mu_0$ and $\sim 2$\% for the scale length. In a few cases (notably for NGC~4578 and NGC~4772; see Figure~\ref{fig:profiles}) we have excluded from the fit those radial ranges where there was clearly a local flux excess, a ``bump'' due to the presence of a particularly strong non-exponential feature, such as a ring. 

The outer limits to our fits have been determined by the confidence limit of the profile brightness, set at 4.94 $\sigma_{\rm sky}$ (see Section~\ref{sec:sky_subtraction}).

\section{The classification scheme}\label{sec:classification}

In this paper we will be using the classification scheme first put forward in \cite{ebp05} and further elaborated in \nocite{pt06}PT06 and \nocite{epb08}Paper~I. This consists of three basic profile types: Type I, a single continuous exponential decline in surface brightness with no change in slope; Type II, a double (or broken) exponential, changing from a shallower inner slope to a steeper outer slope at a defined galactocentric radius; and Type III, similarly a double/broken exponential, but this time changing from a steeper inner slope to a shallower outer slope at a defined galactocentric radius. This nomenclature is based on the original work of \citet{freeman70} who first distinguished Types I and II. 

Type II profiles have a number of subclasses (see \nocite{pt06}PT06 and \nocite{epb08}Paper~I for diagrammatic representations). If a galaxy is barred, then we can distinguish between Type II.i or Type II.o, depending on whether the ``break'', the change in exponential slope, occurs inside (``i'') or outside (``o'') the radius of the end of the bar.  Type II.o can be divided again into two: Type II.o-OLR and Type II.o-CT. It has been suggested (\nocite{epb08}Paper~I) that the former are caused by dynamical effects associated with the Outer Lindblad Resonance (OLR). These breaks are found typically at radii between 2 and 3 bar lengths, which is the zone where outer rings are normally found. Type II.o-CT, ``classical truncations'', do not bear any obvious dynamic relation with any measured features, and we conjecture that they are in some way related to a star formation threshold, and not to the OLR of a bar \citep[e.g.,][]{schaye04,elmegreen06,roskar08a,sanchez-blazquez09}. In certain ambiguous cases, we leave the classification as ``II.o''.  For \textit{unbarred} galaxies, like the vast majority of those in this paper, the subtypes reduce to II-CT or II-AB (the latter being ``asymmetric break'' profiles, which do not appear in our sample; \nocite{pt06}PT06 see them only in Sc -- Sd galaxies).

Finally, Type III profiles can also be divided into subtypes, this time based on the apparent morphology of the outer component.  Type III-d profiles are those where we infer that the outer (shallower) part of the profile is still part of the galaxy's disk, either because we see clear spiral structure in this region or because the observed ellipticity does not change significantly.  In Type III-s profile, on the other hand, we see evidence for the outer component being due to a separate, rounder structure: e.g., an outer ``spheroid''.  The main evidence for this scenario is the presence of isophotes which become systematically rounder beyond the break radius.  For face-on galaxies (e.g., inclination $\lesssim 30 \arcdeg$), the distinction is often difficult or impossible (unless there are positive disk-morphology indications, such as spiral arms at large radii), and so we leave these profiles with a basic ``III'' classification.  We note that in Type III-d profiles the change in slope of the brightness profile between inner and outer components is often abrupt, while for Type III-s profiles the transition at the break radius is gradual (as expected if light from the spheroid is dominating over light from the disk). In a few cases, where the classification was not very clear we left the profile classified as Type III-d(?), or Type III-s(?), with the question mark.

As found by \nocite{pt06}PT06 and \nocite{epb08}Paper~I, there are a minority of profiles which show Type II behaviour in the inner radial range, and Type III behaviour outside. The profiles are composed of three exponential sections, one internal, a second further out with a steeper slope, and a third outermost section with a shallower slope than the second. In Table~\ref{T:M6} of this article we list 5 galaxies identified as Type II + III; these galaxies are: IC~499, NGC~3455, 
NGC~3813, NGC~4369, and NGC~5273.

\begin{deluxetable*}{llrrrrrr}
\tablewidth{0pt}
\tablecaption{Outer-Disk Profile Classifications and Measurements\label{T:M6}} 
\tablecolumns{8}
\tablehead{
\colhead{Galaxy} & \colhead{Profile Type} & \colhead{$h_i$} & \colhead{$h_o$} &
\colhead{$R_{\rm brk}$} & \colhead{$\mu_{0,i}$} & \colhead{$\mu_{0,o}$} & \colhead{$\mu_{\rm brk}$} \\
 &  & \colhead{\arcsec{}} & \colhead{\arcsec{}} & \colhead{\arcsec{}} & & & \\
\colhead{(1)} & \colhead{(2)} & \colhead{(3)} & \colhead{(4)} &
\colhead{(5)} & \colhead{(6)} & \colhead{(7)} & \colhead{(8)}
}
\startdata
IC~356   &    III-d       & 32.9   &  65.7    &     69      &   19.41    & 20.47     &    21.6   \\
IC~499   &    II.o-CT + III-s    & 24.0   &  12.5    &     49      &   20.41    & 18.36     &    22.7   \\
         &                & 12.5   &  50.7    &     80      &   18.36    & 24.44     &    25.3   \\
NGC~278  &    III         & 14.1   &  35.1    &     65      &   18.19    & 21.26     &    23.0   \\
NGC~949  &    III-d       & 31.0   &  117.6   &     138     &   20.36    & 23.95     &    25.0   \\
NGC~972  &    III-s       & 23.2   &  53.5    &     105     &   19.50    & 22.21     &    24.2   \\  %
NGC~1068 &    II.o-OLR    & 125.7  &  73.7    &     190     &   21.37    & 20.31     &    23.0   \\
NGC~1161 &    I           & 92.9   &  \nodata    &   $>$ 327      &   22.50    & \nodata     &  $>$ 26.4   \\
NGC~2300 &    I           & 141.4  &  \nodata    &   $>$ 525      &   22.49    & \nodata     &  $>$ 26.6   \\
NGC~2460 &    II-CT       & 49.7   &  28.4    &     176     &   21.87    & 19.00     &    25.9   \\
NGC~2775 &    III-d(?)    & 26.8   &  61.4    &     90      &   18.71    & 20.75     &    22.1   \\
NGC~2985 &    III-d       & 18.1   &  81.0    &     69      &   18.57    & 21.76     &    21.8   \\ 
NGC~3031 &    II.o-OLR    & 185.8  &  111.4   &     520     &   19.37    & 17.38     &    22.4   \\
NGC~3032 &    I           & 18.9   &  \nodata    &   $>$ 110       &   20.49    & \nodata     &  $>$ 26.9   \\
NGC~3169 &    I           & 53.6   &  \nodata    &   $>$ 360       &   20.50    & \nodata     &  $>$ 27.9   \\
NGC~3245 &    III-s       & 18.9   &  85.3    &     120     &   18.72    & 23.28     &    24.5   \\
NGC~3455 &    II-CT + III-d    & 16.9   &  9.9     &     23      &   19.69    & 18.64     &    21.2   \\
         &                & 9.9    &  21.1    &     40      &   18.64    & 21.00     &    22.9   \\
NGC~3599 &    I           & 23.1   &  \nodata    &   $>$ 168       &   19.97    & \nodata     &  $>$ 27.8   \\
NGC~3604 &    III-d(?)    & 17.2   &  42.8    &     80      &   20.18    & 23.09     &    25.1   \\
NGC~3607 &    I           & 82.1   &  \nodata    &   $>$  360      &   21.45    & \nodata     &  $>$ 26.3   \\
NGC~3619 &    III-d       & 29.4   &  38.4    &     100     &   20.67    & 21.39     &    24.3   \\
NGC~3626 &    I           & 21.6   &  \nodata    &   $>$  142      &   19.20    & \nodata     &  $>$ 26.3   \\
NGC~3675 &    III-s       & 44.9   &  79.2    &     153     &   19.07    & 20.60     &    22.7   \\
NGC~3813 &    II-CT + III-s    & 19.6   &  12.7    &     44      &   18.55    & 17.20     &    21.0   \\
         &                     & 12.7   &  38.1    &     71      &   17.20    & 21.26     &    23.3   \\
NGC~3898 &    III-d       & 30.0   &  59.9    &     111     &   19.54    & 21.53     &    23.3   \\
NGC~3900 &    III-d       & 26.7   &  189.2   &     158     &   20.12    & 25.53     &    25.6   \\
NGC~3998 &    III-d       & 23.7   &  122.0   &     122     &   19.41    & 23.84     &    24.5   \\
NGC~4138 &    III-d       & 15.8   &  21.6    &     54      &   18.32    & 19.33     &    22.0   \\
NGC~4150 &    III-s       & 13.1   &  21.1    &     70      &   18.63    & 20.81     &    24.3   \\
NGC~4223 &    III-s       & 27.3   &  46.1    &     120     &   20.26    & 22.25     &    25.0   \\
NGC~4281 &    III-s       & 19.1   &  92.0    &    91.0     &   18.58    & 22.66     &    20.6   \\
NGC~4369 &    II.o + III    & 24.5   &  14.7    &     50.0      &   20.44    & 19.02     &    22.8   \\
         &                & 14.7   &  67.7    &     114     &   19.02    & 25.61     &    27.3   \\
NGC~4459 &    III-d       & 36.4   &  58.1    &     119     &   19.87    & 21.25     &    23.4   \\
NGC~4578 &    I           & 26.9   &  \nodata    &   $>$  155      &   20.03    & \nodata     &  $>$ 27.1   \\
NGC~4736 &    II.o-OLR    & 541.8  &  134.7   &     373     &   22.6    & 20.33     &    23.5   \\ %
NGC~4750 &    III-d       & 24.2   &  53.9    &     158     &   19.90    & 23.82     &    26.9   \\
NGC~4772 &    I       & 49.7   &  \nodata    &     $>$ 255     &   21.42    & \nodata     &   $>$ 26.9   \\
NGC~4826 &    III-s       & 58.2   &  303.1   &     370     &   18.18    & 23.86     &    24.7   \\ %
NGC~4880 &    II          & 37.8   &  21.9    &     68      &   21.07    & 19.69     &    23.1   \\
NGC~4941 &    I           & 28.5   &  \nodata    &   $>$  232      &   19.56    & \nodata     &  $>$ 28.4   \\
NGC~5273 &    II-CT + III-d    & 22.1   &  15.9    &     58      &   19.83    & 18.75     &    22.7   \\
         &                & 15.9   &  23.0    &     83      &   18.75    & 20.48     &    24.4   \\
NGC~5485 &    I           & 113.5  &  \nodata    &   $>$  400      &   23.17    & \nodata     &  $>$ 27.9   \\
NGC~5520 &    I           & 12.4   &  \nodata    &   $>$  82      &   19.34    & \nodata     &  $>$ 26.9   \\
NGC~6340 &    III         & 26.6   &  50.5    &     110     &   19.87    & 22.00     &    24.2   \\
NGC~7217 &    III         & 24.3   &  63.1    &      79     &   18.76    & 20.99     &    21.8   \\
NGC~7457 &    III-d       & 22.1   &  33.2    &     42      &   19.01    & 19.74     &    21.0   \\
UGC~3580 &    I           & 27.3   &  \nodata    &   $>$  158      &   20.98    & \nodata     &  $>$ 27.4   \\
UGC~4599 &    III-d       & 42.2   &  64.1    &     121     &   24.05\tablenotemark{a}    & 25.07\tablenotemark{a}   &   27.2\tablenotemark{a}   \\
\enddata

\tablenotetext{a}{$B$-band values}

\tablecomments{Classification and measurements of disks surface brightness
profiles.  Surface brightnesses are observed values, and have not been
corrected for Galactic extinction, inclination, or redshift. Note that Type~I
profiles by definition have no ``outer'' part and have only lower limits for
the break radius and $\mu_{\rm brk}$.  For galaxies with composite profiles
(IC~499, NGC~3455, NGC~3813, NGC~4369, and NGC~5273), we list values for both
the inner zone (Type~II) and the outer zone (Type~III). Columns: (1) Galaxy
name; (2)  Profile classification as explained in text; (3) and (4) Scale
length for the inner and outer exponential fits, respectively; (5) Position of
break point on the profile; (6) and (7) Central $R$-band surface brightness in
mag arcsec$^{-2}$ for the inner and outer exponential fits, respectively; (8)
Surface brightness at the break point.} \end{deluxetable*}

\begin{deluxetable*}{lrrrrrrr}
\tablewidth{0pt}
\tablecaption{Disk Parameters in Physical and Relative Units\label{T:M7}} 
\tablecolumns{8}
\tablehead{
\colhead{Galaxy} & \colhead{$h_i$} & \colhead{$h_i$} & \colhead{$h_o$} &
\colhead{$h_o$} & \colhead{$R_{\rm brk}$} & \colhead{$R_{\rm brk}$} & \colhead{$R_{\rm brk}$} \\
 & \colhead{($R_{25}$)} & \colhead{(kpc)} & \colhead{($R_{25}$)} & \colhead{(kpc)} & \colhead{($R_{25}$)} & \colhead{(kpc)} & \colhead{($h_i$)} \\
\colhead{(1)} & \colhead{(2)} & \colhead{(3)} & \colhead{(4)} &
\colhead{(5)} & \colhead{(6)} & \colhead{(7)} & \colhead{(8)}
}
\startdata
IC 356    &  0.21  &  2.41  &  0.42  &  4.81  &  0.44  &  5.05  &  2.10     \\
IC 499   &  0.38  &  3.35  &  0.20  &  1.75  &  0.78  &  6.85  &  2.04     \\
         &  0.20  &  1.75  &  0.81  &  7.08  &  1.28  &  11.18  &  6.40     \\
NGC 278  &  0.22  &  0.75  &  0.56  &  1.87  &  1.04  &  3.47  &  4.63     \\
NGC 949  &  0.43  &  1.65  &  1.63  &  6.25  &  1.92  &  7.34  &  4.45     \\
NGC 972  &  0.23  &  2.44  &  0.54  &  5.62  &  1.06  &  11.03  &  4.53     \\
NGC 1068  &  0.59  &  8.68  &  0.35  &  5.09  &  0.89  &  13.11  &  1.51     \\ %
NGC 1161  &  1.10  &  12.54  &  \nodata  &  \nodata  &  \nodata  &  \nodata  &  $>$ 3.52     \\
NGC 2300  &  1.67  &  20.06  &  \nodata  &  \nodata  &  \nodata  &  \nodata  &  $>$ 3.72     \\
NGC 2460  &  0.67  &  5.35  &  0.39  &  3.06  &  2.38  &  18.91  &  3.54     \\
NGC 2775  &  0.21  &  2.33  &  0.48  &  5.34  &  0.70  &  7.83  &  3.36     \\
NGC 2985  &  0.13  &  1.86  &  0.59  &  8.28  &  0.50  &  7.01  &  3.78     \\
NGC 3031  &  0.23  &  3.27  &  0.14  &  1.96  &  0.64  &  9.15  &  2.80     \\
NGC 3032  &  0.32  &  1.96  &  \nodata  &  \nodata  &  \nodata  &  \nodata  &  $>$ 5.82     \\
NGC 3169  &  0.41  &  4.28  &  \nodata  &  \nodata  &  \nodata  &  \nodata  &  $>$ 6.71     \\
NGC 3245  &  0.19  &  1.86  &  0.88  &  8.40  &  1.24  &  11.82  &  6.35     \\
  NGC 3455  &  0.23  &  1.30  &  0.13  &  0.76  &  0.32  &  1.78  &  1.38     \\
          &  0.13  &  0.76  &  0.29  &  1.62  &  0.55  &  3.09  &  4.10     \\
NGC 3599  &  0.29  &  2.22  &  \nodata  &  \nodata  &  \nodata  &  \nodata  &  $>$ 7.27     \\
NGC 3604  &  0.27  &  1.79  &  0.68  &  4.45  &  1.28  &  8.31  &  4.65     \\
NGC 3607  &  0.56  &  8.83  &  \nodata  &  \nodata  &  \nodata  &  \nodata  &  $>$ 4.38     \\
NGC 3619  &  0.36  &  3.39  &  0.48  &  4.43  &  1.24  &  11.55  &  3.40     \\
NGC 3626  &  0.27  &  2.04  &  \nodata  &  \nodata  &  \nodata  &  \nodata  &  $>$ 6.57     \\
NGC 3675  &  0.25  &  2.81  &  0.45  &  4.95  &  0.87  &  9.59  &  3.42     \\
  NGC 3813  &  0.29  &  2.08  &  0.19  &  1.35  &  0.65  &  4.60  &  2.23     \\
          &  0.19  &  1.35  &  0.57  &  4.05  &  1.05  &  7.52  &  5.57     \\
NGC 3898  &  0.23  &  2.75  &  0.46  &  5.49  &  0.85  &  10.16  &  3.70     \\
NGC 3900  &  0.28  &  3.35  &  1.99  &  23.75  &  1.67  &  19.84  &  5.92     \\
NGC 3998  &  0.29  &  1.58  &  1.51  &  8.13  &  1.51  &  8.13  &  5.14     \\
NGC 4138  &  0.21  &  1.03  &  0.28  &  1.40  &  0.70  &  3.52  &  3.41     \\
NGC 4150  &  0.19  &  0.85  &  0.30  &  1.37  &  1.00  &  4.54  &  5.35     \\
NGC 4223  &  0.45  &  2.18  &  0.75  &  3.69  &  1.96  &  9.60  &  4.40     \\
NGC 4281  &  0.21  &  2.20  &  1.02  & 10.60  &  1.0   & 10.48  &  4.76     \\
NGC 4369  &  0.39  &  1.98  &  0.23  &  1.19  &  0.80  &  4.03  &  2.04     \\
          &  0.23  &  1.19  &  1.08  &  5.46  &  1.91  &  9.68  &  8.16     \\
NGC 4459  &  0.34  &  2.84  &  0.55  &  4.52  &  1.12  &  9.27  &  3.27     \\
NGC 4578  &  0.27  &  2.12  &  \nodata  &  \nodata  &  \nodata  &  \nodata  &  $>$ 5.76    \\
NGC 4736  &  1.61  &  13.29  &  0.40  &  3.30  &  1.11  &  9.14  &  0.69     \\
NGC 4750  &  0.40  &  2.98  &  0.88  &  6.63  &  2.59  &  19.48  &  6.55     \\
NGC 4772  &  0.49  &  3.50  &  \nodata  &  \nodata  &  \nodata  &  \nodata  &  $>$ 5.13   \\
NGC 4826  &  0.19  &  2.05  &  1.01  &  10.69  &  1.23  &  13.05  &  6.36     \\
NGC 4880  &  0.40  &  3.61  &  0.23  &  2.09  &  0.72  &  6.49  &  1.80     \\
NGC 4941  &  0.26  &  2.07  &  \nodata  &  \nodata  &  \nodata  &  \nodata  &  $>$ 8.14    \\
NGC 5273  &  0.27  &  1.72  &  0.19  &  1.24  &  0.70  &  4.52  &  2.62     \\
          &  0.19  &  1.24  &  0.28  &  1.79  &  1.00  &  6.47  &  5.22     \\
NGC 5485  &  1.61  &  13.89  &  \nodata  &  \nodata  &  \nodata  &  \nodata  &  $>$ 3.52     \\
NGC 5520  &  0.21  &  1.71  &  \nodata  &  \nodata  &  \nodata  &  \nodata  &  $>$ 6.61     \\
NGC 6340  &  0.27  &  2.60  &  0.52  &  4.94  &  1.13  &  10.77  &  4.14     \\
NGC 7217  &  0.21  &  1.76  &  0.54  &  4.56  &  0.68  &  5.71  &  3.25     \\
NGC 7457  &  0.17  &  1.38  &  0.26  &  2.07  &  0.33  &  2.62  &  1.90     \\
UGC 3580  &  0.27  &  2.55  &  \nodata  &  \nodata  &  \nodata  &  \nodata  &  $>$ 5.79    \\
UGC 4599  &  0.71  &  5.77  &  1.07  &  8.76  &  2.02  &  16.54  &  2.87     \\
\enddata

\tablecomments{As for Table~\ref{T:M6}, but now listing outer-disk parameters in alternate units. Columns: (1) Galaxy name; (2) and (3) Scale lengths for the inner exponential fits in units of $R_{25}$ and kpc, respectively; (4) and (5) Scale lengths for the outer exponential fits in units of $R_{25}$ and kpc, respectively; (6), (7), and (8) Radius of the break point on the profile in terms of $R_{25}$, kpc, and the inner scale length.}
\end{deluxetable*}

\section{Bars in ``Unbarred'' Galaxies}\label{sec:bars}

As noted above (Section~\ref{sec:sample}), the sample studied in this paper was intended to consist solely of \textit{unbarred} galaxies; this includes some galaxies classified as barred in the RC3 which were, for various reasons, judged to be unbarred \citep[see][]{erwin03a,erwin05}.  Nonetheless, it is by now well accepted that large optical surveys such as RC3 sometimes miss bars, especially when galaxies are so dusty that the bars are more easily seen in infrared light \citep[e.g.,][]{eskridge00}.  Accordingly, we have carefully examined each of the galaxies in our sample (making use of near-IR images when possible, as well as prior studies of each galaxy in the literature) for ``hidden'' bars.  This examination has turned up a total of ten galaxies classified as unbarred which \textit{do} host bars.

We also include in this section, 2 galaxies which were in fact classified as barred by RC3, but for which arguments had previously been made for the absence of bars, resulting in their exclusion from \nocite{epb08}Paper~I and inclusion in this sample.  In some cases, we agree with the arguments (e.g., NGC~3032 and NGC~3455), but in the case of NGC~278 and NGC~4941 we now argue that these galaxies \textit{do} have bars.

\begin{figure*}
  \centering
  \includegraphics[width=5.0in]{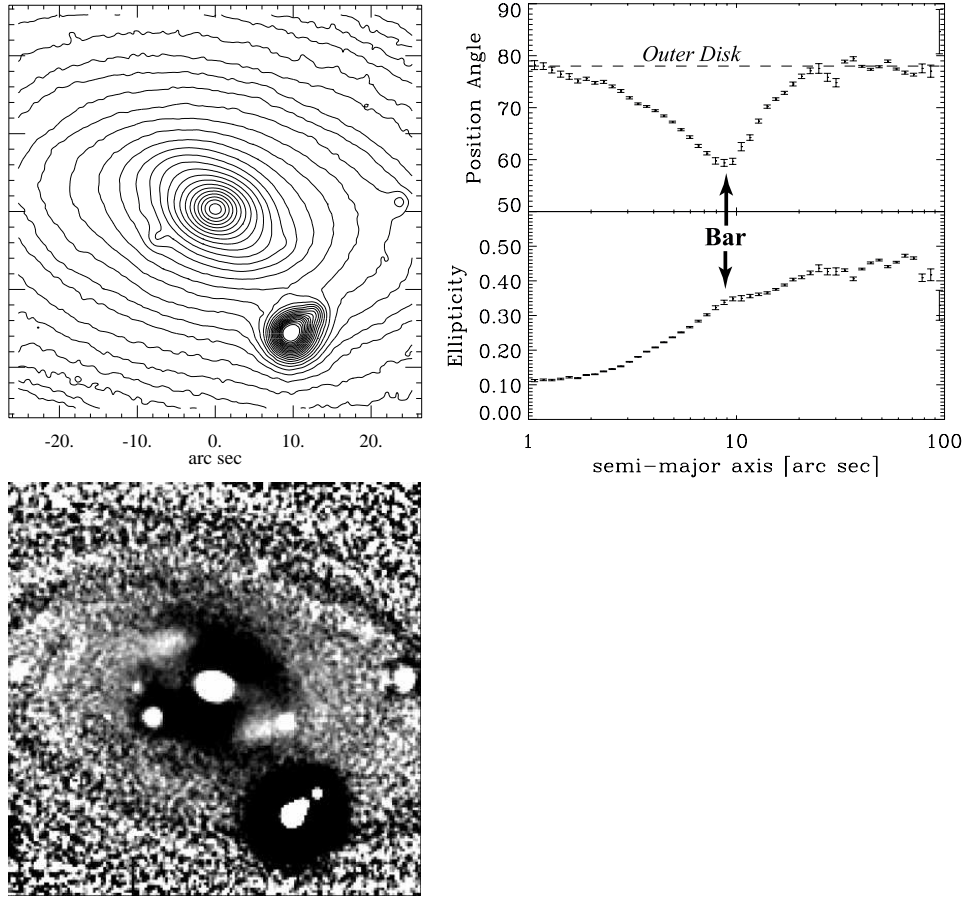}
  \caption{Evidence for a bar in the Sa galaxy IC~499. Upper left panel: $R$-band isophotes, logarithmically spaced.  Upper right panel: ellipse fits to the same image (position angle of outer disk is indicated by horizontal dashed line); the bar shows up primarily as a sharp twist in the position angles.  Lower left panel: unsharp mask of the $R$-band isophotes (same scale as upper left panel).} \label{fig:ic499}
\end{figure*}
\begin{figure*}
  \centering
  \includegraphics[width=5.0in]{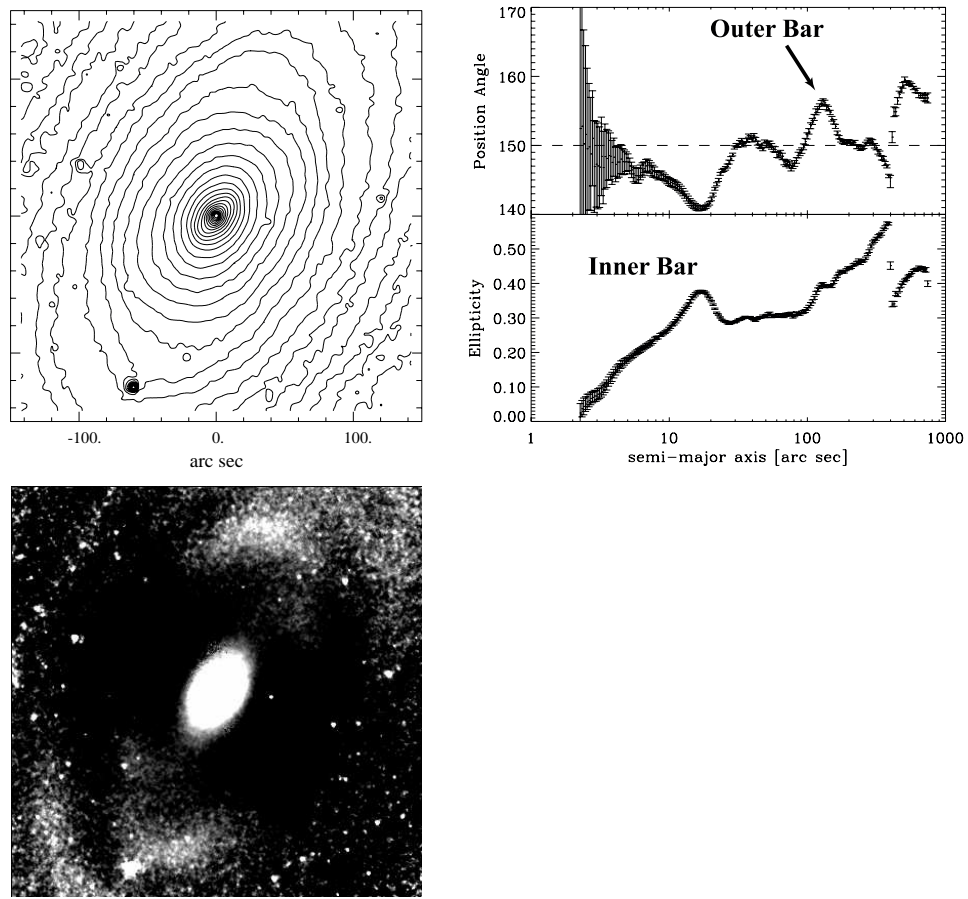}
  \caption{As for Figure~\ref{fig:ic499}, but now showing evidence for a large-scale (``outer'') bar in the SAab galaxy NGC~3031 (M81), using the Spitzer IRAC1 image from \citet{dale09}.} \label{fig:n3031}
\end{figure*}
\begin{figure*}
  \centering
  \includegraphics[width=5.0in]{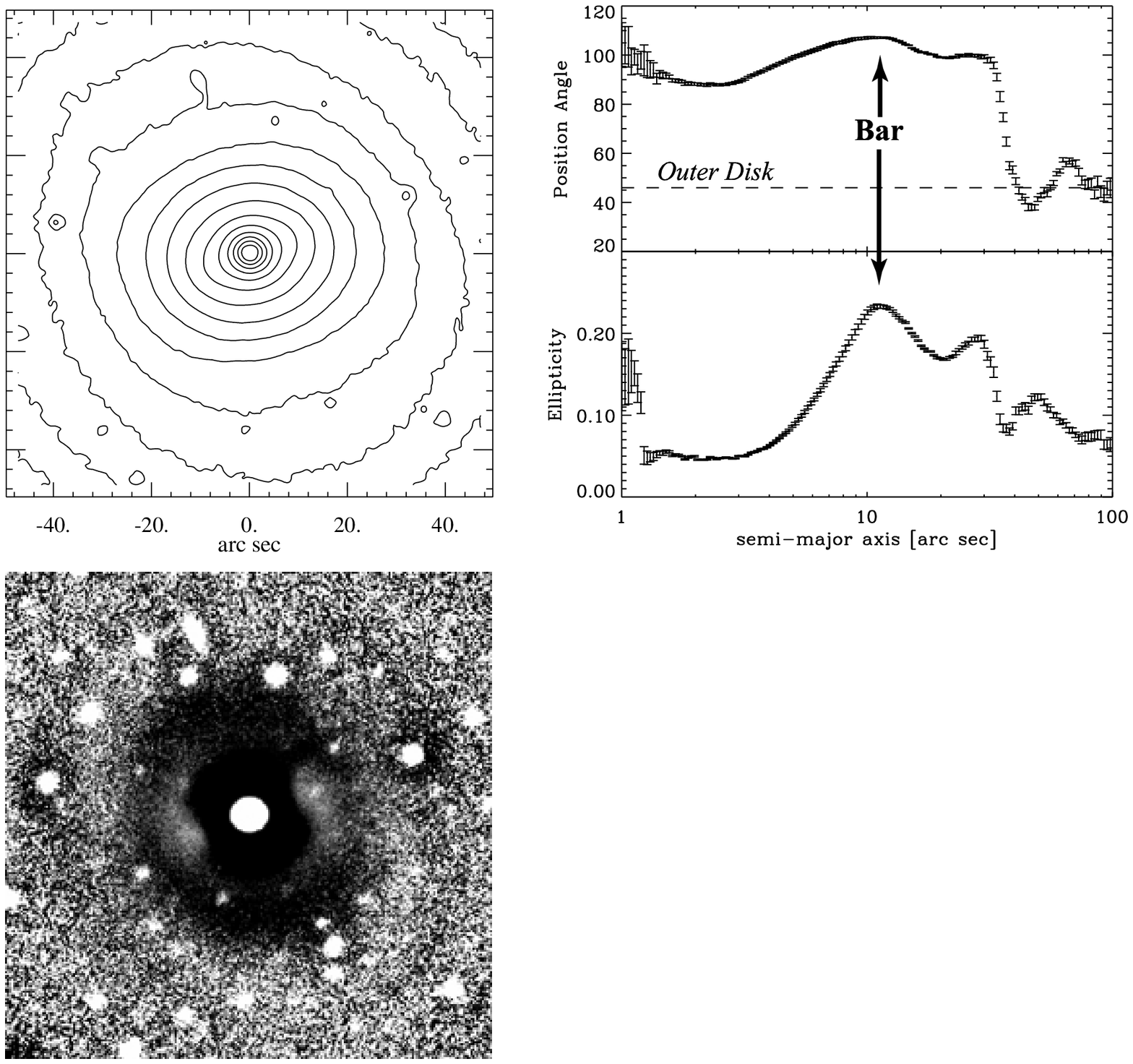}
  \caption{As for Figure~\ref{fig:ic499}, but now showing evidence for a bar in the SA0 galaxy NGC~3599.} \label{fig:n3599}
\end{figure*}
\begin{figure*}
  \centering
  \includegraphics[width=5.0in]{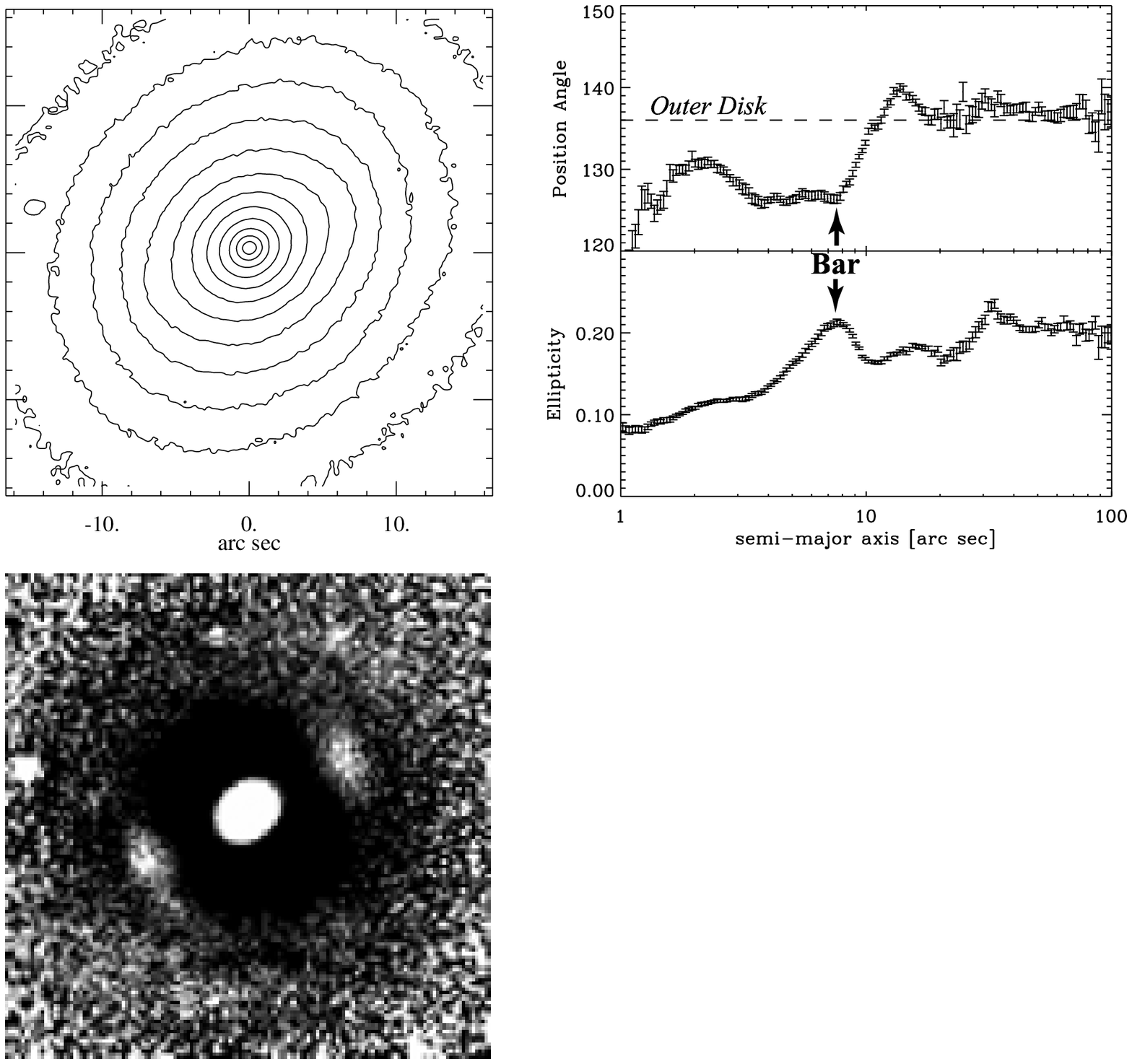}
  \caption{As for Figure~\ref{fig:ic499}, but now showing evidence for a bar in the SA0 galaxy NGC~3998.} \label{fig:n3998}
\end{figure*}

Undetected bars in optically ``unbarred'' galaxies fall into three
general categories.  The first are small- or medium-sized bars which are
relatively weak and which have no accompanying rings or dust lanes;
these may, especially in relatively featureless S0 galaxies, be
difficult to distinguish with photographic plate data, 
due to the combination of poor resolution and contrast.
Figures~\ref{fig:ic499}, \ref{fig:n3031}, \ref{fig:n3599}, and
\ref{fig:n3998} show bars in IC~499, NGC~3031, NGC~3599, and NGC~3998. 
In all cases, the bars are relatively small, but show up in ellipse fits
and unsharp masks (the characteristic feature of many bars in unsharp
masks is caused by the sharp drop in surface brightness at the bar ends;
see, e.g., \nocite{erwin03a}Erwin \& Sparke 2003). Additional evidence
for the (outer) bar in NGC~3031 comes from an isophote morphology which
indicates a vertically thickened bar seen at an intermediate angle (see
Erwin \& Debattista, in prep, for more
details).

\begin{figure*}
  \centering
  \includegraphics[width=5.0in]{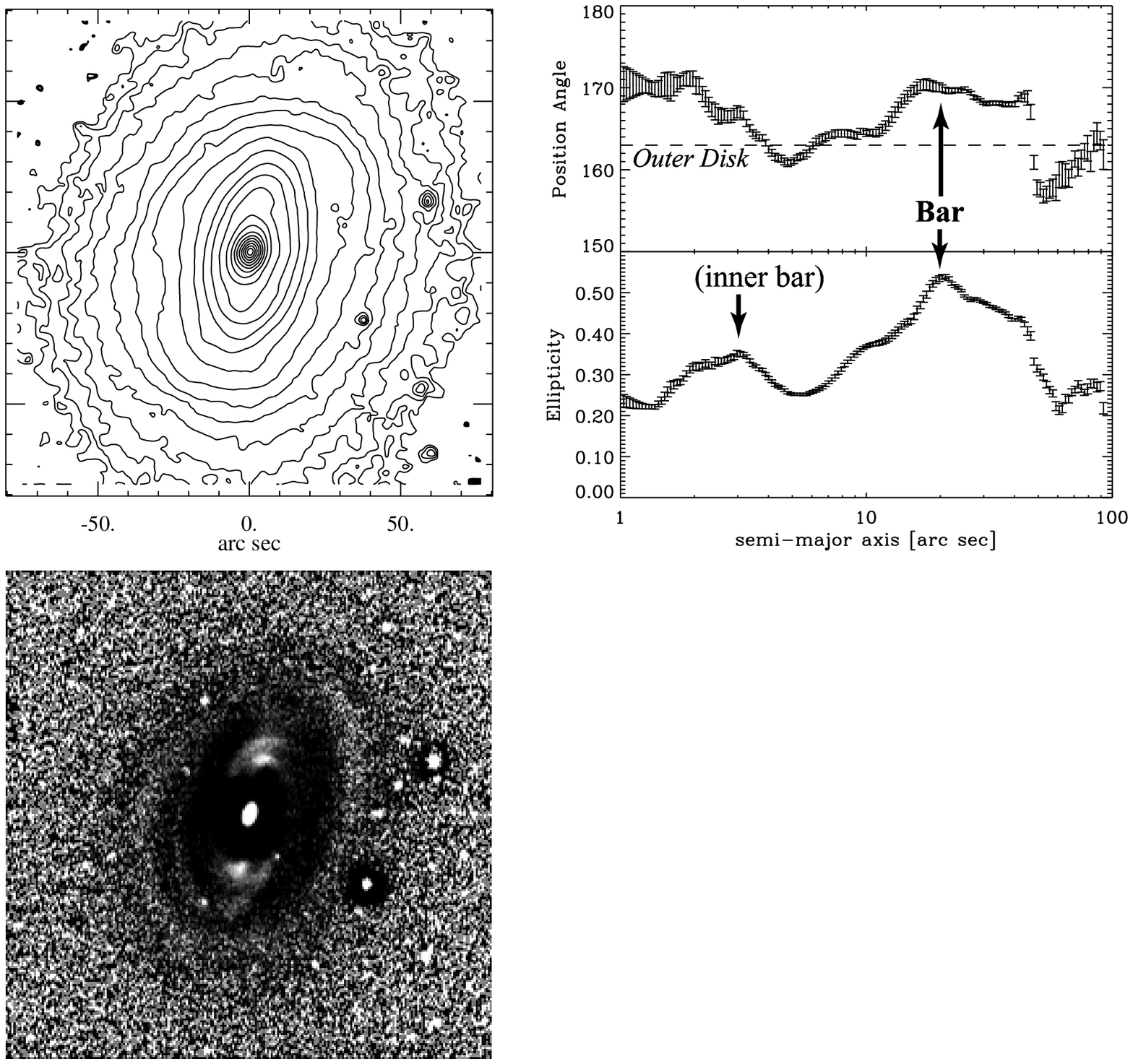}
  \caption{As for Figure~\ref{fig:ic499}, but now showing evidence for a bar in the SA0 galaxy NGC~3626, using $J$-band image from \citet{mollenhoff01}.} \label{fig:n3626}  
\end{figure*}

The second class of hidden bar consists of small- or medium-sized bars
which have been obscured by dust; these are then best seen with
near-infrared images.  The most famous example of this in our sample is
NGC~1068, whose (inner) bar was first pointed out by \citet{scoville88}.
Other examples include: NGC~3031, where \citet{elmegreen95} used near-IR
images to point out the existence of a small, weak bar in the center of
this galaxy; NGC~3626, where \citet{laurikainen05} found both a
large-scale bar and evidence for a nuclear bar using $K$-band images
(see Figure~\ref{fig:n3626}); NGC~4369, in which a small, strong bar was
reported by \citet{knapen00}, based on the near-IR images of
\citet{peletier99}; NGC~4736 \citep{shaw93,mollenhoff95}; and NGC~4750
\citep{laine02}.

The third class of hidden bars are those in galaxies where the bar is so \textit{large}, and the disk outside the bar so low in contrast and surface brightness, that the bar has been \textit{mistaken} for the outer disk; it then requires deeper imaging to bring out the true outer disk and show that the bar is more elliptical and (often) misaligned with respect to the outermost isophotes.  A discussion of this phenomenon for the case of NGC~5248 (not part of our sample) is given by \citet{jogee02}.  When the bar is relatively weak (i.e., not highly elliptical), these are sometimes referred to as ``oval disks'' \citep{kormendy79}. NGC~1068 and NGC~4736 are also examples of this phenomenon (where here we refer to their large, outer bars, not the dust-obscured inner bars discussed in the previous paragraph).


As noted above, NGC~1068, NGC~3031, NGC~4736, and NGC~4750 all have
small bars which can be seen clearly only in near-IR images; such bars
are sometimes referred to as ``nuclear'' bars, though in the case of 
NGC~1068 the bar in question is almost three kpc in diameter.  These
galaxies \textit{also} have much larger, weaker bars, which in some
cases (e.g., on shallow images) can be mistaken for the outer disk (a
partial exception is NGC~3031, which has obvious spiral structure
extending well outside the weak outer bar).  Thus, they are actually
\textit{double}-barred galaxies \citep[NGC~3626 may also fall into this
category; see][]{laurikainen05}.  \citet{erwin04} summarizes the
evidence for large-scale bars in NGC~1068 and NGC~4736.

\begin{figure*}
  \centering
  \includegraphics[width=5.0in]{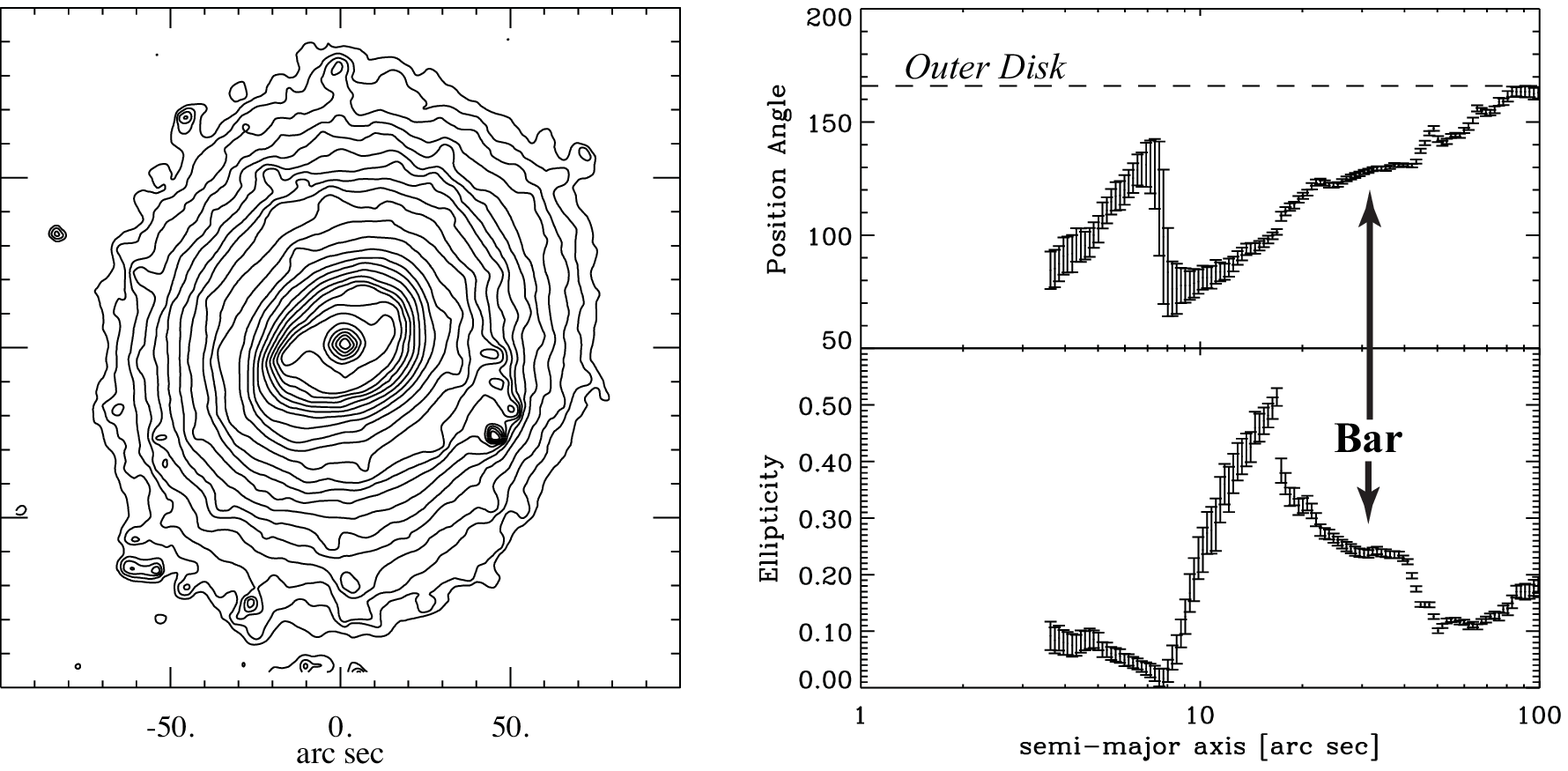}
  \caption{Spitzer IRAC2 contours (left, logarithmically spaced) and ellipse fits (right) for NGC~4750.  The weak, oval bar is at a position angle of $\sim 125\arcdeg$; the ellipticity peak at $a \sim 15\arcsec$ is due to spiral arms inside.} \label{fig:n4750}
\end{figure*}

Figure~\ref{fig:n4750} shows isophotes and ellipse fits for an archival
$3.6\mu$m Spitzer image of NGC~4750 (Program ID 40349, PI Giovanni
Fazio).  There is a weak oval structure with $a \sim 30\arcsec$, a
position angle of $\sim 125\arcdeg$, and ellipticity $\sim 0.3$.
\textit{Inside} this oval is a pair of spiral arms; these  arms are what
\citet{laine02} identified as an ``outer bar'' with a
semi-major axis of 14\arcsec. This interior spiral structure is an
indication that the oval is not a conventional strong bar. Nonetheless,
the oval is clearly misaligned with respect to the outer disk, and more
elliptical, so we consider it a bar.

\begin{figure}
  \centering
  \includegraphics[width=2.8in]{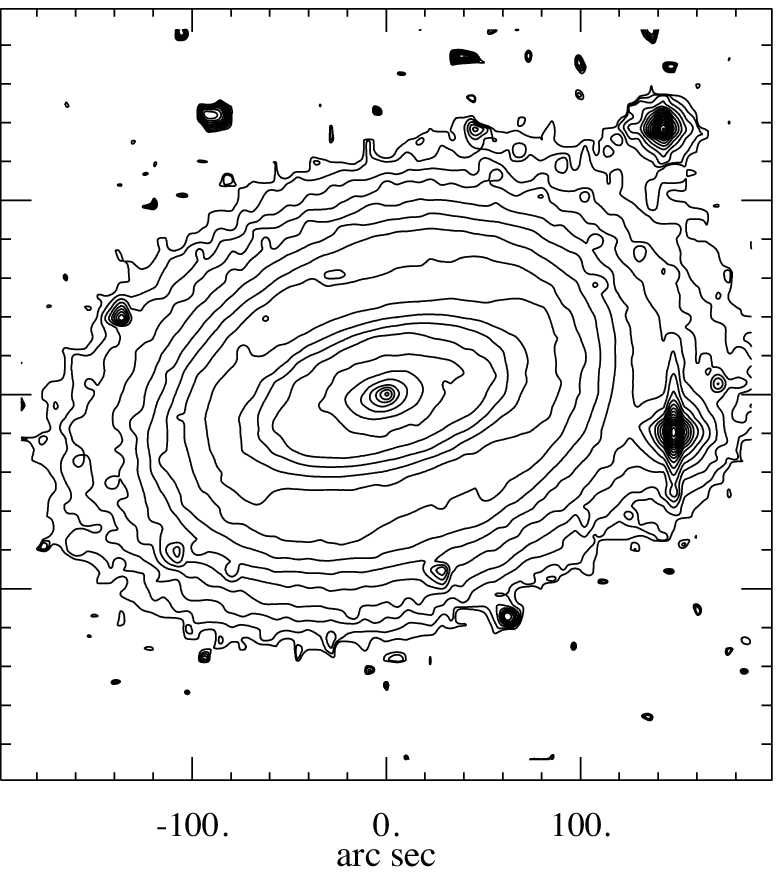}
  \caption{$R$-band contours for NGC~4941 ($\mu_{R} = 27$ to 17, in steps of 0.5 mag arcsec$^{-1}$). The large, oval bar is surrounded by an outer ring ($a \sim 100\arcsec{}$); the outer disk shows up as the outer set of elliptical isophotes.} \label{fig:n4941}
\end{figure}

A similar case is NGC~4941. A nuclear bar with semi-major axis $\sim 250$
pc was seen in near-infrared images of this galaxy by
\citet{greusard00}, but those authors also argued that this was the only
bar in the system, and that the RC3 classification of SAB was erroneous;
this led to our excluding this galaxy from the barred-galaxy sample of
\nocite{epb08}Paper~I, since that paper was supposed to include only galaxies 
with large-scale bars. However, \citet{kormendy82b} suggested that this
galaxy have a (large-scale) ``oval disk'' like those in NGC~1068 and 
NGC~4736.  Our analysis suggests that NGC~4941 is in fact a double-bar
system analogous to those two galaxies. Figure~\ref{fig:n4941} shows
large-scale isophotes for this galaxy.

\begin{figure}
  \centering
  \includegraphics[width=2.8in]{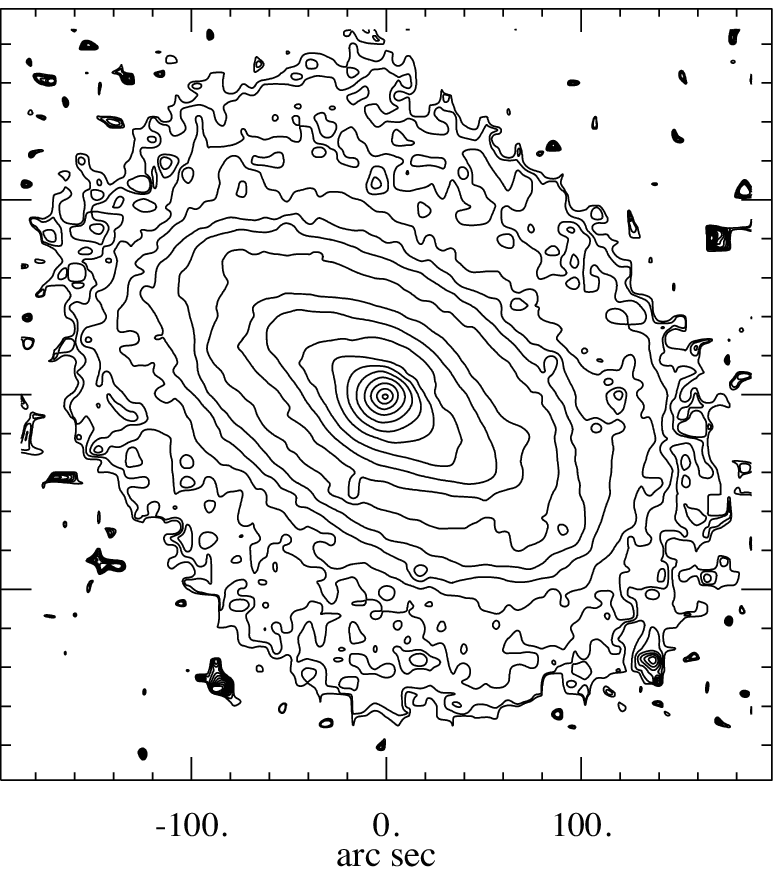}
  \caption{$R$-band contours for NGC~4772 ($\mu_{R} = 26$ to 17, in steps of 0.5 mag arcsec$^{-1}$). The large, oval bar is surrounded by fainter, rounder isophotes forming what may be a partial outer ring.} \label{fig:n4772}
\end{figure}

The final and most ambiguous example is NGC~4772 (see Figure~\ref{fig:n4772}). 
This galaxy appears,
at first glance to consist of a luminous, round bulge embedded in a
highly inclined, dusty disk, whose axis ratio is given by RC3.  The
latter structure extends to $r \approx 140\arcsec{}$ along its major
axis.  However, outside this elliptical structure there is a much
fainter and rounder structure with slightly boxy isophotes, with a small
twist with respect to the inner isophotes.  This can be seen in the
SDSS images, and was noted previously in $B$-band images by
\citet{haynes00}, who pointed out that these outer isophotes coincided
with a rounder outer ring in the \hi{} distribution.  They also noted a
misalignment between the inner stellar and \ha{} kinematics, suggestive
of ``a misaligned disk or bar''.  The $H$-band image of
\citet{eskridge02} shows that the ``disk'' region isophotes become
slightly rectangular-ended (at $r \sim 80\arcsec{}$), as is often the
case for bars in early-type disks \citep{athanassoula90}.

The overall appearance is quite similar to several other early-type
spirals with large bars and faint outer rings, such as NGC~4941
(Figure~\ref{fig:n4941}) and NGC~5377 \citep[see, e.g.,][]{erwin03a}. 
Combined with the evidence 
described by \citet{haynes00}, this leads us to identify the bright,
elliptical ``disk'' of NGC~4772 as a very large, weak bar (or ``oval
disk''); we note that \citet{eskridge02} made a similar classification.
Unfortunately, the faintness of the outer isophotes makes it
somewhat difficult to determine the true orientation of the disk, though
the stellar kinematics of Haynes et al.\ do suggest that the major axis
is close to 147\arcdeg{}.

We define the bar's ellipticity and $\amax$ using ellipse fits to the
$H$-band image of \citet{eskridge02}, with the upper limit on its size
being set by measurements of the star-forming ring (which we tentatively
identify as an inner ring) using the GALEX NUV image of \citet{gdp07}. 
The isophotes remain highly elliptical outside this region (out to $a
\sim 130\arcsec{}$); it is unclear whether this should be considered
part of the bar.

Table~\ref{T:bars} presents bar measurements for the galaxies discussed in this Section, using the approach of \citet{erwin05}.  For double-barred galaxies, we list just the large-scale ``primary'' bars.


\begin{deluxetable*}{lrrcrr}
\tablewidth{0pt}
\tablecaption{Bar Parameters\label{T:bars}} 
\tablecolumns{6}
\tablehead{
\colhead{Galaxy} & \colhead{Bar PA (\arcdeg{})} & \colhead{\amax (\arcsec{})} & \colhead{\amin/\aten} &
\colhead{\lbar (\arcsec{})} & \colhead{\emax}}
\startdata
IC~499   &  49 &  8.8 &  \nodata/\nodata &   10 &  0.35 \\ 
NGC~278  & 105 &   15 &   19/16 &   16 &  0.25 \\ 
NGC~1068 &  12 &   54 &   75/89 &   75 &  0.24 \\ 
NGC~3031 & 157 &  134 &  \nodata/\nodata &  215 &  0.40 \\ 
NGC~3599 & 106 &   11 &   21/16 &   16 &  0.23 \\ 
NGC~3626 & 172 &   20 &   35/46 &   35 &  0.53 \\ 
NGC~3998 & 126 &  7.8 &   11/11 &   11 &  0.21 \\ 
NGC~4369 & 156 &  4.5 &  \nodata/10 &   10 &  0.65 \\ 
NGC~4736 &  90 &  125 &  170/\nodata &  170 &  0.23 \\ 
NGC~4750 & 127 &   33 &   45/44 &   44 &  0.24 \\ 
NGC~4772 & 146 &   70 &  \nodata/\nodata &   80 &  0.53 \\ 
NGC~4941 &  17 &   68 &  \nodata/\nodata &   95 &  0.41 \\ 
\enddata
\tablecomments{Parameters for bars in galaxies previously classified as unbarred.  \amax{} is the semi-major axis of maximum isophotal ellipticity, closest to the bar end, and is a lower limit on bar size. \amin{} is the semi-major axis of minimum ellipticity outside the bar end, while \aten{} is the semi-major axis at which the position angle of fitted ellipses varies by more than 10\arcdeg{} from the bar's position angle.  \lbar{} is the adopted upper limit on bar size. (The minimum and maximum of these values are plotted in Fig. \ref{fig:profiles}.) \emax{} is the maximum isophotal ellipticity within the bar.  For double-barred galaxies (NGC~1068, 3031, 4736, and 4941, and possibly NGC~3626), we list measurements for the outer bar only; values for NGC~1068 and 4736 are taken from \citet{erwin04}.}
\end{deluxetable*}

How common, then, are ``hidden'' or otherwise unrecognized bars?  If we take the combined S0--Sb sample of \nocite{epb08}Paper~I and this paper, we have a total of 38 galaxies classified as optically unbarred (SA), and another 5 galaxies with no bar classification (S).  Of these, we find bars in 9 of the former ($24 \pm 7$\%) and 1 of the latter ($\sim 20$\%).  Conversely, there appear to be no bars in two\footnote{NGC~3032 and NGC~3455} of the 76 optically barred (SB or SAB) galaxies in the combined sample; three more optically barred galaxies were excluded due to their being polar ring or merger systems (NGC~2146, NGC~2655, and NGC~2685).  Roughly speaking, then, we can argue that $<$ 5\% of optically barred galaxies are mis-classified, while $\sim 20$\% of optically unbarred galaxies prove, on closer inspection, to host bars.

These misclassification fractions are lower than those found by \citet{eskridge00} from their analysis of $H$-band images of a sample of spiral galaxies, where they found that half of the optically unbarred spiral galaxies showed the presence of a bar. Note that galaxies studied by Eskridge et al.\ are later types --- predominantly Sbc and Sc --- and therefore dustier.

\section{Results and Discussion}\label{sec:discussion}

\subsection{Trends with Hubble Type and Global Frequencies of Disk Profile Types}\label{sec:hubble-trends}

By combining the sample studied in this paper with that previously studied in \nocite{epb08}Paper~I \textit{and} the late-type spirals (Sbc--Sdm) studied by \nocite{pt06}PT06, we can, for the first time, see how profile types vary along the entire Hubble sequence of disk galaxies, and how common the different profiles are in the general population of disk galaxies.

\begin{figure}
  \centering
  \includegraphics[width=3.5in]{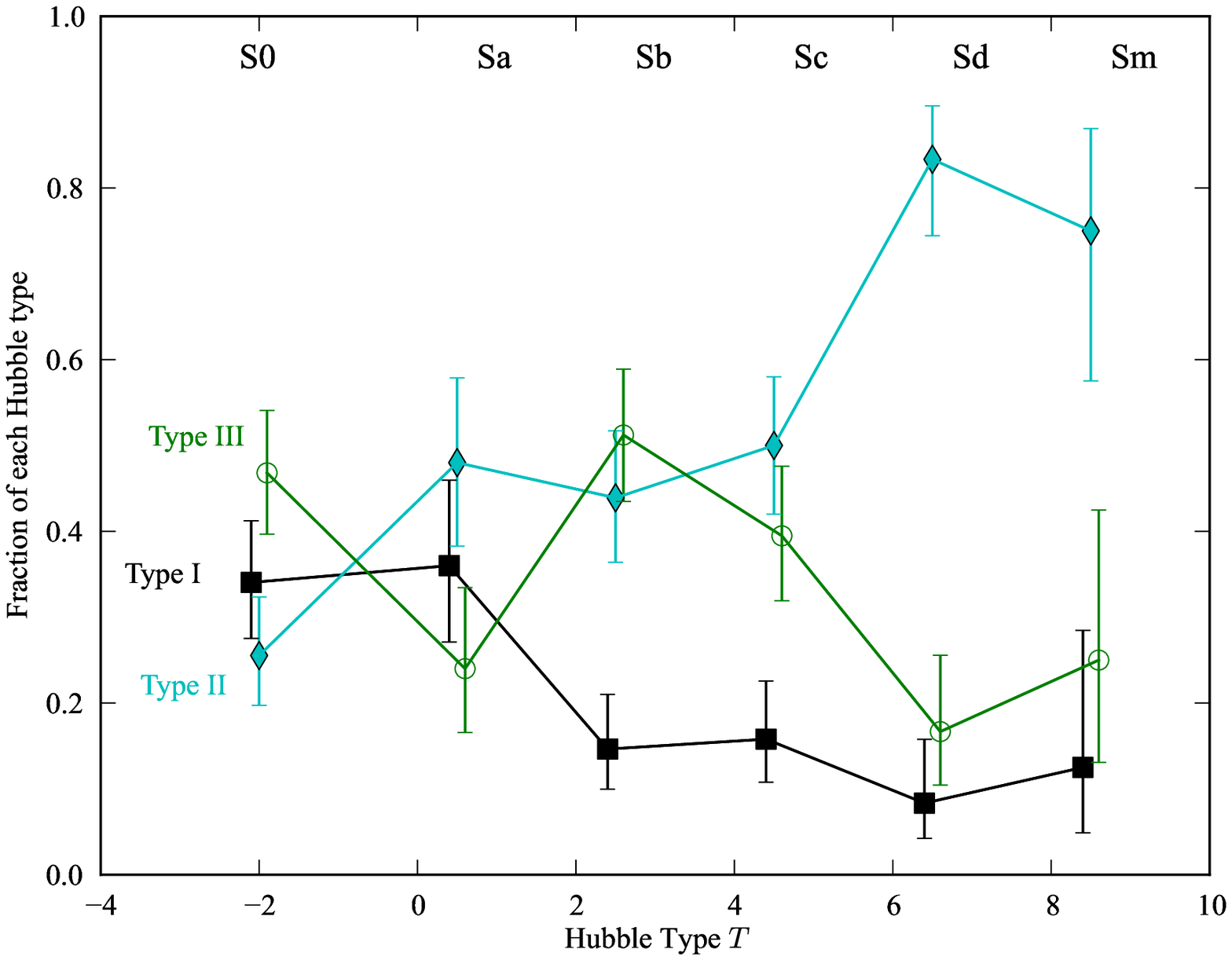}
  \caption{Frequencies of the basic outer-disk profile types along the Hubble sequence.  This plot uses data from the combined sample, including the early-type barred galaxies from Paper~I and the late-type galaxies from PT06.  Error bars represent 68\% confidence intervals, based on the Wilson confidence interval for binomial statistics (see text).} \label{fig:fractions}
\end{figure}

Figure~\ref{fig:fractions} plots the fractions of the three main profile types
as a function of Hubble type for the combined S0--Sdm samples. The bins for early-type disks (S0, S0/a--Sa, Sab--Sb) are taken from \nocite{epb08}{Paper~I} and this paper; profiles for late-type disks (Sbc--Sc, Scd--Sd, Sdm--Sm) are taken from \nocite{pt06}PT06. (To avoid mixing different sample selections into the same bin due the partial overlap at Sb between the early- and late-type samples, we excluded the Sb galaxies in the PT06 sample.) 
The plot uses classifications for a total of 183 galaxies: 113 early-type disk galaxies from this paper and \nocite{epb08}Paper~I, and 70 Sbc and later-type galaxies from \citet{pt06}. For Type~II we count all possible subtypes (e.g., II.i, II.o, II-AB, etc.) and all ``composite-profile'' galaxies (i.e., II+III profiles). Similarly, the Type~III fractions count all subtypes (III-d, III-s, and plain III) \textit{and} all composite-profile galaxies. Composite-profile galaxies are thus counted twice, so the totals in each bin may be greater than 1.
The error bars are 68\% confidence limits derived from the so-called ``Wilson'' binomial confidence interval \citep{wilson27}, rather than the commonly used but inaccurate Gaussian (``Wald'') approximation; see \cite{brown01} for a discussion of these issues.

Two fairly clear trends emerge from this figure. The first is the relatively small fraction of Type~II profiles in early type disks, and the dramatic increase in later Hubble types: while only $\sim 25$\% of S0 galaxies have Type~II profiles, the fraction is $\sim 80$\% for the very latest spirals.  The second trend is a clear, though less dramatic, change in the frequency of Type~I profiles, which are most common in early-type disks and least common in late-type spirals.

We can also combine the various studies to estimate the \textit{global} fractions of different profile types, although we must be careful in doing so, since the early and late-type samples were constructed differently and have different degrees of completeness.  The early-type sample studied in \nocite{epb08}Paper~I and this paper was constructed to cover field S0--Sb (and Virgo S0) galaxies from the UGC with $\delta \geq -10\arcdeg$, axis ratios $\leq 2.0$, major-axis diameters $D_{25} \geq 2.0\arcmin$, and redshifts $\leq 2000 $\kms; it is essentially complete within these constraints. (Note that Virgo Cluster spirals were excluded from the sample.)  The late-type sample of \nocite{pt06}PT06 was constructed to cover Sb--Sdm galaxies with the same axis ratio restriction, but with a redshift limit of $V \leq 3250$ \kms, Galactic latitudes $|b_{\rm II}| > 20\arcdeg$, and $M_B < -18.4$. Because their data source was Data Release 2 (DR2) of SDSS, the final sample is incomplete: only $\sim 15$\% of the galaxies found in the HyperLeda database meeting those criteria actually had DR2 images, and some of the images were not usable (e.g., galaxy too close to the edge of an image).  Clearly, we cannot simply add up all the profile numbers across both samples and expect to get fully significant results.

Instead, we correct the individual profile counts within each sample for selection effects, using a common notional parent sample, which we base on that of \nocite{pt06}PT06: all disk galaxies ($-3.5 < T < 8.5$) in the HyperLeda database with axis ratios $< 2.0$ ($\log r_{25} < 0.301$), $V \leq 3250$ \kms, $|b_{\rm II}| > 20\arcdeg$, and $M_B < -18.4$.  We then remove those galaxies from \nocite{epb08}Paper~I and this paper which do not meet the parent-sample criteria (a total of 12 galaxies) and compute the completeness for the two observed samples (S0's and early-type spirals: 102/591; late-type spirals: 70/537); the inverses of these completeness fractions are used to scale the observed counts of profile types.

From this we obtain our estimates for the global frequencies of disk-profile types: $21 \pm 3$\% Type I, $50 \pm 4$\% Type II, and $38 \pm 4$\% Type III. \footnote{Uncertainties are estimated by rescaling all corrected counts so that the totals add up to the number of observed galaxies (172), and then calculating the Wilson confidence intervals.} \textit{Composite} Type II+III profiles (e.g., IC~499, NGC~3455, NGC~3813, NGC~4399, and NGC~5273 in this paper) are here counted as both Type II and Type III; such profiles account for $8 \pm 2$\% of the total.  The frequency of ``pure'' Type III profiles is $29 \pm 3$\%.

\subsection{Comparing Parameters for Different Disk Profile Types}

\begin{figure*}
  \centering
  \includegraphics[width=6.0in]{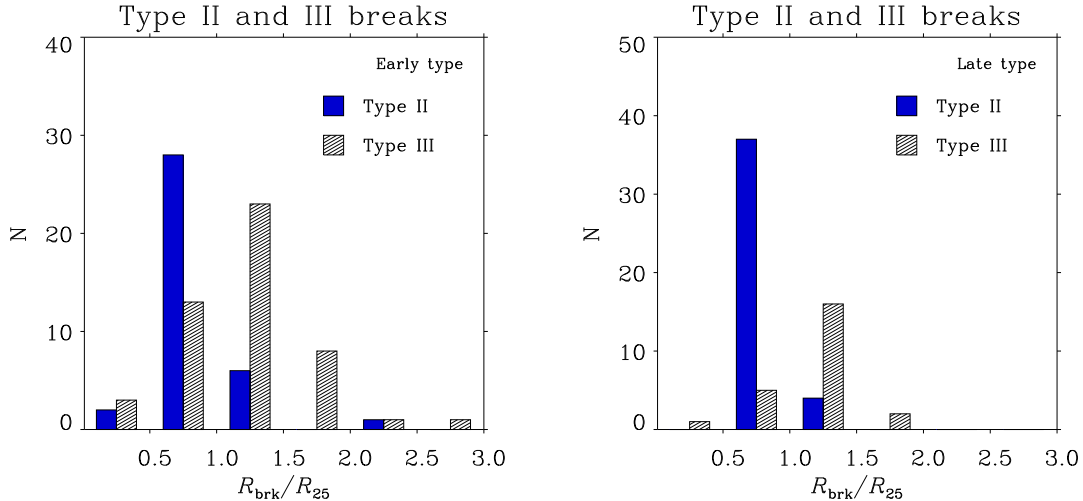}
  \caption{Histograms of break radii for Type II and III profiles, in units of $R_{25}$, including barred and unbarred galaxies. The left-hand panel uses data from this paper and Paper~I (S0--Sb galaxies), while the right-hand panel uses data from PT06 (Sbc--Sdm galaxies). The blue and hashed-black bars represent all the galaxies of each type in a given bin range, but are presented with the intervals between them for greater clarity, since we are plotting two histograms together.} \label{fig:breaks}
\end{figure*}

In Figure~\ref{fig:breaks}, we plot histograms for the position of the
break radius for Type II and III profiles, in units of $R_{25}$.  This
includes all galaxies from this study and \nocite{epb08}Paper~I (grouped
together as ``early-type'' disks, left panel), along with separate plots
for the late-type sample of \nocite{pt06}PT06 (right panel).  We exclude
the (rare) Type II.i profiles from this comparison.  Although the median
sizes of the break radii are not drastically different (e.g., $0.79 \,
R_{25}$ for Type II versus $1.02 \, R_{25}$ for Type III in the
early-type sample), the Type III break radii clearly have a broader
spread and are weighted toward larger values.  A Kolmogorov-Smirnov
(K-S) test confirms that the differences between the two profile types
are significant ($P = 0.0022$ for the early-type sample and $P = 2.1
\times 10^{-7}$ for the late-type sample).  Given that breaks at very
large radii are harder to detect --- since they are likely to occur at
or beyond the reliability limit of our surface photometry --- it is
possible that we are underestimating the number of Type III breaks with
radii $\gtrsim 2 R_{25}$. It is also worth noting that $R_{25}$ is a
surface-brightness limit, and so changes in the surface-brightness
profile can in principle change $R_{25}$ --- if those changes happen
interior to the $\mu_{B} = 25$ level. The sense of this change would
actually be the \textit{opposite} of what we see: a truncation would
bring $R_{25}$ closer in, and thus make $R_{\rm brk}/R_{25}$
\textit{larger}, while an antitruncation would push $R_{25}$ further
out, making $R_{\rm brk}/R_{25}$ \textit{smaller}.  If this effect is
operating in our sample (and it cannot be operating for those galaxies
with $R_{\rm brk}/R_{25} > 1$), it means that we are, if anything,
\textit{underestimating} the difference between Type II and Type III
break radii.

\begin{figure*}
  \centering
  \includegraphics[height=8.0in]{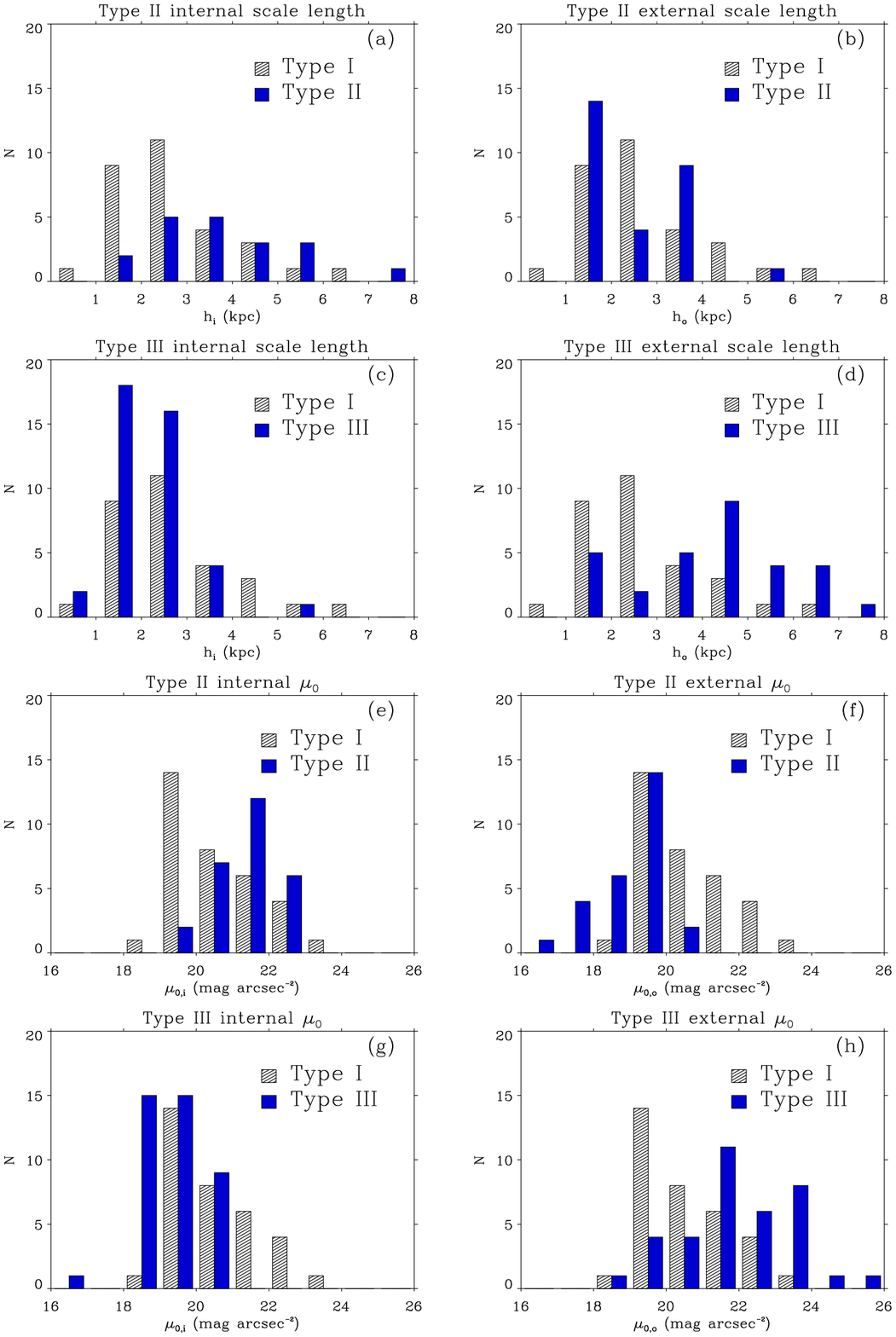}
  \caption{Distributions of scale length $h$, in units of kpc, and of extrapolated central surface brightness $\mu_0$.  We show these separately for the inner and outer exponential fits where there is a truncation (Type II) or an antitruncation (Type III); these are compared in each panel with $h$ or $\mu_{0}$ for the single exponentials of Type I profiles (hashed-black bars).  This plot uses data from our sample data and includes the early-type barred galaxies from Paper~I. As in Fig. \ref{fig:breaks}, the blue and hashed-black bars represent all the galaxies of each type in the given bin range, but are presented with the intervals between them for greater clarity.} \label{fig:h_mu_early}
\end{figure*}

\begin{figure*}
  \centering
  \includegraphics[height=8.0in]{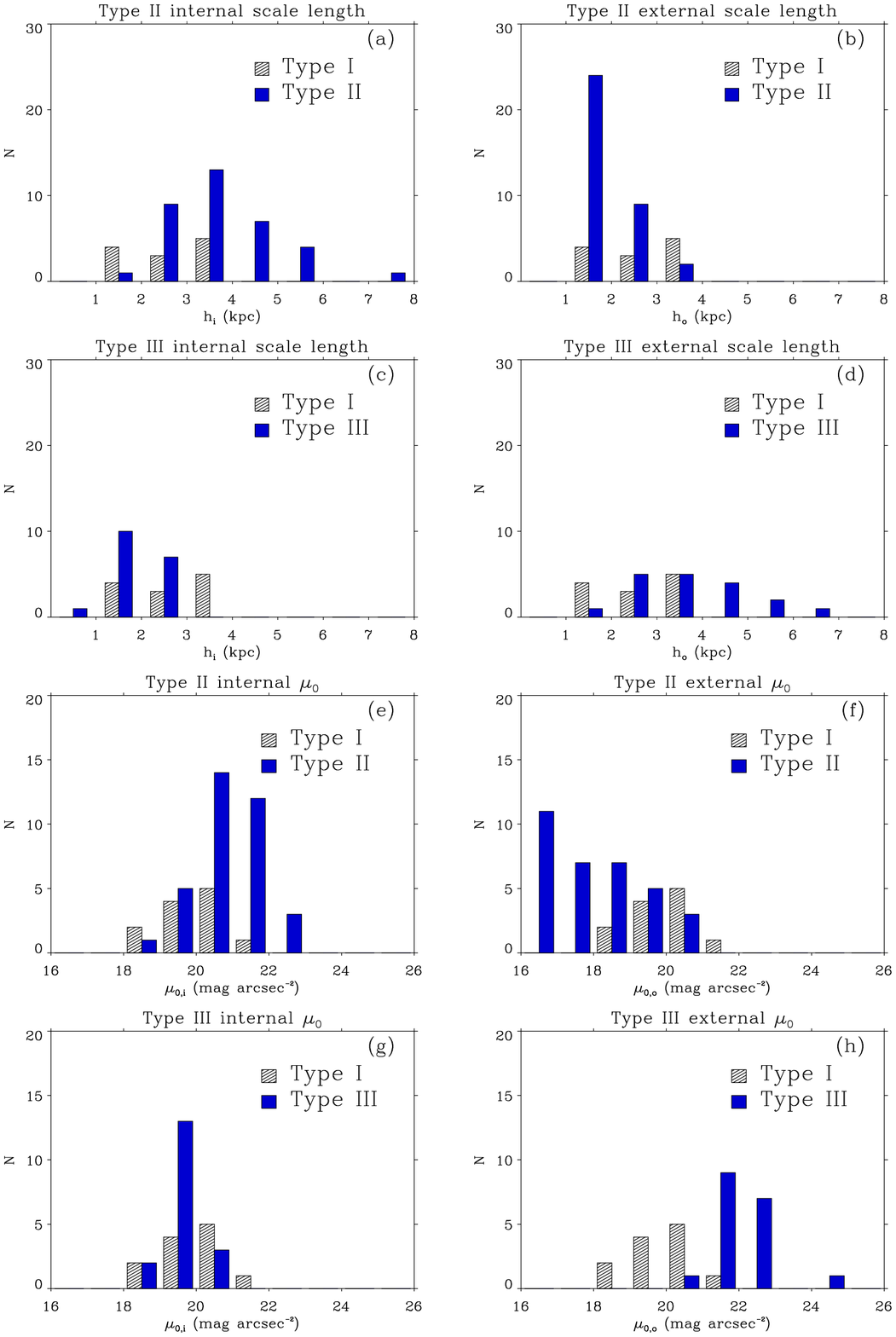}
  \caption{As in Figure~\ref{fig:h_mu_early}, but this plot uses only data from the late-type galaxies (Sbc -- Sbm) from PT06.} \label{fig:h_mu_late}
\end{figure*}

Though truncated and antitruncated profiles differ, by definition, from the simpler Type I profiles, there may still be some underlying similarities. We could hypothesize, for example, that the inner regions of Type II and III profiles (that is, the profile interior to the break radius) are fundamentally similar to Type I profiles, so that the only real difference is in the region outside the break.  Alternatively, there could be physical similarities between the \textit{outer} parts of Type II or III profiles and Type I profiles --- as well as the possibility that none of the sub-regions are similar. In Figures~\ref{fig:h_mu_early} and \ref{fig:h_mu_late} we plot histograms of absolute disk scale lengths $h$, along with histograms of the extrapolated central surface brightnesses $\mu_0$.  These are plotted separately for the inner and outer components of Type II and III profiles, along with histograms of the same parameters for the single exponentials of Type I profiles.  As with Figure~\ref{fig:breaks}, we plot the early-type and late-type samples separately.

The only potential similarity that we can identify is between Type I profiles and the \textit{inner} parts of Type II profiles. The latter do tend to have longer scale lengths and fainter $\mu_0$ values, but K-S tests do not rule the null hypothesis of the same parent population ($P = 0.11$ and 0.10 for scale lengths of early- and late-type disks, respectively; $P = 0.18$ and 0.14 for $\mu_0$ values). So scenarios in which (at least some) Type II profiles are merely truncations of what would otherwise be Type I profiles are still plausible.  Not surprisingly, the \textit{outer} components of Type II profiles have significantly shorter scale lengths and brighter $\mu_0$ values, and do \textit{not} match Type I profiles ($P = 0.0012$ and 0.0042 for early- and late-type scale lengths; $P = 1.1 \times 10^{-8}$ and 0.00025 for $\mu_0$).

Analysis of the late-type spirals in the PT06 sample by \citet{bakos08} showed that Type II profiles tend to have $g - r$ color profiles that are bluest at the break \citep[also seen for $u - g$ colors in galaxies out to $z \sim 1$;][]{azzollini08}. Bakos et al.\ showed that the stellar surface-density profiles of late-type truncations were smoothed out --- sometimes to the point of looking like Type I profiles. The difference between the stellar and the surface-brightness profiles is apparently due to the relative youth of stars near the break radius, with the stellar populations located away from the break (on both sides) being predominantly older. The difference we observe between the parameters of Type I and (many) Type II profiles is at least broadly consistent with this: an excess of light near the break due to younger stellar populations will tend to flatten the inner part of a Type II profile and push its estimated $\mu_0$ to fainter values.

However, we should note that there do exist individual Type II profiles --- in particular, some of the Type II.o-OLR profiles in \nocite{epb08}Paper~I --- where the inner zone is either extremely flat (very large scale length) or simply not exponential (the latter values are obviously not included in the histograms and statistical tests), so not all Type II profiles can be characterized this way; see Erwin et al. (in prep) for more discussion of the differences in, and possible origins of, Type II profiles.

On the other hand, we can probably rule out common parent populations for Type I profiles and both parts of Type III profiles. The inner components of Type III profiles have shorter scale lengths and significantly brighter $\mu_0$ values ($P = 0.032$ and 0.03 for early- and late-type scale lengths; $P = 0.0037$ and 0.00092 for $\mu_0$).  
This suggests that scenarios where Type III profiles are formed merely by adding or redistributing stars at large radii to a pre-existing Type I profile may not explain all such cases. This includes the minor-merger simulations of \citet{younger07}, where the inner scale lengths of their Type III profiles were very similar to the initial (single-exponential) scale length of the primary galaxy.
The scale lengths of Type III \textit{outer} components do not differ drastically from those of Type I profiles ($P = 0.015$ and 0.074 for early- and late-types), but the $\mu_0$ values do: Type III outer components have $\mu_0$ values $\sim 1$ mag arcsec$^{-2}$ brighter ($P = 0.0062$ and $6.1 \times 10^{-5}$).

\begin{figure}
  \centering
  \includegraphics[width=3.4in]{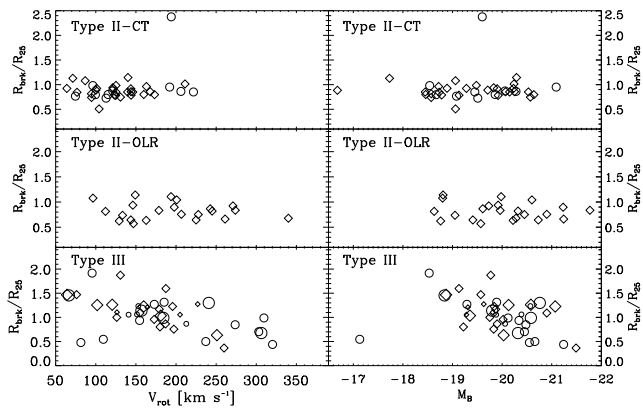}
  \caption{Plots of break radius, in units of $R_{25}$, against rotation velocity (left column) and absolute magnitude (right column), separately for Type II-CT, for Type II-OLR, and for Type III profiles. This plot uses data from the combined sample, including the early-type barred galaxies from Paper~I and the late-type galaxies from PT06. Symbols are coded as follows: unbarred galaxies are circles, barred galaxies are diamonds; in the third row, small-size symbols are Type III-s profiles, medium-size symbols are III-d, and larger symbols are non-specific Type III profiles.} \label{fig:breaks_velocity}
\end{figure}

Finally, in Figure~\ref{fig:breaks_velocity} we plot the break radii (in units of $R_{25}$) against the deprojected rotation velocities (left hand panels) and against absolute blue magnitude (right hand panels); we do this separately for Type II-CT, Type II-OLR, and Type III profiles.  
Rotation velocities are derived primarily from HyperLeda observed gas rotation velocities (deprojected using our values for the inclination) for spiral galaxies\footnote{Since gas in S0  galaxies is frequently misaligned with respect to the stellar disk \citep[e.g.,][]{kuijken96,davis11}, we do not attempt to do this for S0 galaxies.}, supplemented by stellar circular velocities from \citet{neistein99} for some of the S0 galaxies from \nocite{epb08}Paper~I.
Overall trends within each profile type are weak or absent, except possibly for Type III profiles, where the break radius tends to be smaller for higher rotation velocities and luminosities.  Type II-CT tend to occur in galaxies with low values of rotation velocity, compared to the Type II-OLR and Type III profiles; the median values of the rotation velocity for each profile type are 125.0, 193.7, and 195.7 km s$^{-1}$, respectively.  This is probably due the higher frequency of classical truncations (II-CT) in late-type spirals, which tend to have lower rotation velocities.  As \nocite{pt06}PT06 show (their Figure~11), the majority of Type II profiles in Sc and later Hubble types are II-CT (amounting to $\sim 40$\% of all Sc and later galaxies); by contrast, only $\sim 5$\% of the S0--Sb galaxies in \nocite{epb08}Paper~I and this paper are II-CT.

\section{Summary of the Main Conclusions}\label{sec:conclusions}

In this paper, we have presented azimuthally averaged $R$-band surface-brightness profiles for 47 early-type (S0--Sb) disk galaxies.  These galaxies, mostly unbarred, complement and complete a sample of 66 barred S0--Sb galaxies presented in \nocite{epb08}Paper~I.  The profiles are derived from a variety of images, about half of them obtained with the Isaac Newton Telescope's Wide Field Camera and most of the rest from the Sloan Digital Sky Survey. The profiles are classified following a scheme first sketched out in \citet{erwin05} (based in part on the original scheme of \nocite{freeman70}Freeman 1970) and elaborated in \nocite{pt06}PT06 and Paper~I: Type~I (profile is a single exponential), Type~II (profile steepens  at large radii, including so-called ``truncations''), and Type~III (``antitruncations'', where the profile becomes shallower at large radii).  A small subset of the profiles are composite ``Type II+III'' systems, where the inner part of the disk has a Type~II shape, with an additional, shallower profile at the largest radii.

Although the sample presented in this paper was intended to be purely unbarred galaxies, careful analysis of the images turned up bars of various sizes and strengths in a total of twelve of the galaxies.  This implies that $\sim 20$\% of optically unbarred S0--Sb galaxies (that is, galaxies with RC3 classifications of S or SA) are actually barred at some level.

The combination of this paper's galaxies and those in Paper~I forms a set of 113 S0--Sb disk galaxies with profile classifications. By combining this with the 70 Sbc--Sdm galaxies with profile classifications in \nocite{pt06}PT06, we can make the first general statements about how common the different disk-profile types are, and their dependence on Hubble type in the local universe.  The traditional idea that galaxy disks are either all pure exponential \textit{or} all radially truncated, already weakened as a result of previous related studies on well resolved nearby galaxy images (\nocite{epb08}Paper~I; \nocite{erwin05}Erwin et al.\ 2005; \nocite{pt06}PT06), is clearly not valid. The global frequencies we find are 21\% Type I, 50\% Type II, and 38\% Type III; 8\% of the galaxies are composite Type II+III profiles. Including barred and unbarred galaxies, we find strong trends with Hubble type: Type II profiles increase from only $\sim 25$\% of S0 galaxies to $\sim 80$\% of the latest-type spirals, while Type I profiles decrease in frequency from early-type disks ($\sim 30$\% of galaxies) to late-type spirals (only $\sim 10$\% of galaxies).  Comparisons of exponential fits to the different profiles suggests a possible similarity in slope and projected central surface brightness for Type I profiles and the \textit{inner} parts of Type II profiles.  However, neither the inner nor outer parts of Type III profiles resemble Type I profiles, which suggests that antitruncations are not simply excess light added at large radii to Type I profiles.

\acknowledgments 

We would like to thank David Wilman and Louis Abramson for useful and
interesting discussions. L.G. thanks DGAPA-UNAM, M\'exico, for support
provided through a PASPA fellowship, and the Instituto de Astrof\'isica
de Canarias, where this work was advanced, for hospitality during his 
PhD thesis.  P.E. was supported by DFG Priority Program 1177 (``Witnesses 
of Cosmic History:  Formation and evolution of black holes, galaxies 
and their environment'').  This work was also supported by grants No.\ 
AYA2004-08251-CO2-01 and AYA2007-67625-CO2-01 from the Spanish Ministry 
of Science and P3/86 of the Instituto de Astrof\'isica de Canarias.  
P.E. acknowledges the hospitality of the IAC during April, 2007.
Also we are very grateful to the referee for a very careful reading of 
the original version of the article and comments which have helped us to improve it.

This research is based in part on observations made with the Isaac
Newton Telescope, which is operated on the island of La Palma by the
Isaac Newton Group in the Spanish Observatorio del Roque de los
Muchachos of the Instituto de Astrof\'isica de Canarias. This paper also
makes use of data obtained from the Isaac Newton Group Archive, which is
maintained as part of the CASU Astronomical Data Centre at the Institute
of Astronomy, Cambridge.

Funding for the creation and distribution of the SDSS Archive has been
provided by the Alfred P. Sloan Foundation, the Participating
Institutions, the National Aeronautics and Space Administration, the
National Science Foundation, the U.S. Department of Energy, the
Japanese Monbukagakusho, and the Max Planck Society.  The SDSS Web
site is http://www.sdss.org/.

The SDSS is managed by the Astrophysical Research Consortium (ARC) for
the Participating Institutions.  The Participating Institutions are
The University of Chicago, Fermilab, the Institute for Advanced Study,
the Japan Participation Group, The Johns Hopkins University, the
Korean Scientist Group, Los Alamos National Laboratory, the
Max-Planck-Institute for Astronomy (MPIA), the Max-Planck-Institute
for Astrophysics (MPA), New Mexico State University, University of
Pittsburgh, University of Portsmouth, Princeton University, the United
States Naval Observatory, and the University of Washington.

This research also made use of the Lyon-Meudon Extragalactic Database
(LEDA; http: //leda.univ-lyon1.fr) and the NASA/IPAC Extragalactic
Database (NED); the latter is operated by the Jet Propulsion Laboratory,
California Institute of Technology, under contract with the National
Aeronautics and Space Administration.

\appendix
\section{Surface Brightness Profiles and specific notes for each galaxy}\label{sec:profiles}
	
In Figure~\ref{fig:profiles} we present the surface brightness profiles of all the galaxies analyzed in this paper. They are the azimuthally averaged values of the surface brightness made using the ellipse fits discussed above, using calibrated $R$-band images, 
plotted against position along the semimajor axis in arcsec and also in kpc. In each plot the level of 4.94 $\sigma_{\rm sky}$ is shown as a horizontal dashed line, and $R_{25}$, taken from the NED (Nasa/IPAC Extragalactic Database), is shown with a vertical arrow. In the few galaxies where a bar was found to be present the bar length is indicated with vertical dotted lines, and any ring features have their radii noted with vertical dashed-dotted lines. Exponentials were fitted to sections of the surface-brightness profile where the log plots are linear. The boundaries of these sections were defined by eye; the $R_{brk}$ was defined as the point were the two fits intersect. See Section~4.5 of \nocite{epb08}Paper~I for a detailed description. The exponential fits are shown as one or more dashed straight lines superposed on the profiles. We now include specific notes on each galaxy.  In the following notes when we refer to the bulge of the galaxy we mean the ``photometric bulge''; this is the region where the surface brightness profile becomes brighter at small radii than the inward projection of the disk (or the inward projection of the inner part of the disk in the case of Type II and III profiles). For each galaxy, along with the profile type, we list the RC3 classification, the same as shown in Table \ref{T:BasicGalaxyData}.

{\bf IC~356 (III-d; SA(s)ab pec)}: 
The disk of this galaxy dominates the light at  $r
> 30\arcsec$ (2 kpc) from the center of the galaxy.  The inner disk clearly shows
spiral structure with multiple arms and dust lanes, and the bulge also
shows dust lanes.  Tightly wrapped spirals are clearly visible outside
the antitruncation break, and the ellipticity is roughly constant out to
$r \sim 300\arcsec$, so this is a clear III-d profile.  The profile is
similar in some respects to that of NGC~4612 (Paper~I), though there is no
bar in IC 356.

{\bf IC~499 (II.o-CT + III-s; Sa)}: 
Though this galaxy is not classified as
barred, we find an inner bar of length $\sim 10\arcsec$ (1.5 kpc) (Section~\ref{sec:bars}). The profile has a clear Type~II break at $r \sim 48.5\arcsec$ (6.8 kpc); since this is more than four times the bar radius, we consider this a ``classical truncation'' rather than an OLR break.  Further outside, we find an antitruncation at a radius of $\sim 93\arcsec$ (13 kpc).  The ellipticity declines from this point outwards, which leads to our classification of Type III-s.

{\bf NGC~278 (III; SAB(rs)b)}: 
This galaxy appears to host a very weak bar;
evidence for this is discussed in Section~\ref{sec:bars} \citep[see
also][]{garrido03}. The wiggles 
between about 11$\arcsec{}$ and 30$\arcsec{}$ come from a dust ring of 
radius 11$\arcsec$ (0.6 kpc) and tightly 
wound arms forming a pseudoring at 20$\arcsec$ (1 kpc) radius. There is no visible
structure in the outer zone, and the galaxy is so close to face-on that
we cannot discriminate between disk and spheroid morphologies for the
outermost light.

{\bf NGC~949 (III-d; SA(rs)b)}: 
The bulge shows strong streaks of dust, and is really bright out to 15.6$\arcsec$ (830 pc). Further out the disk is smooth and shows no real evidence of spiral structure. The isophotes beyond the break radius appear to have approximately the same shape as those inside (though there may be a slight position-angle twist), so we consider this a III-d profile.
As the ellipse fits show a roughly constant ellipticity inwards as far as some 10 arcseconds from the center, 
it is probable that nominal photometric bulge (r $>$ 0 arcsec) is in fact a ``disky pseudobulge'' 
\citep[see][]{kormendy04,erwin03b}.

{\bf NGC~972 (III-s; Sab)}: 
This has a complex bulge with an obvious high dust content, irregularly distributed. The spiral form can just be discerned in the disk. Beyond 43$\arcsec$ (4.5 kpc) the disk is smooth, and gives way to two lobes further out (more visible to the North) which may be arm fragments, or an external pseudoring at 90$\arcsec$ (9.5 kpc). It is at this point where the antitruncation begins. The ellipticity starts to fall systematically from the break, which accounts for our classification as Type III-s. 

{\bf NGC~1068 (II.o-OLR; (R)SA(rs)b)}: 
This is the closest Seyfert 2 galaxy. Although the RC3 classification is (R)SA(rs)b, it is in fact double-barred. \citet{erwin04} summarizes the evidence for both bars, including the inner bar first detected in the NIR by \citet{scoville88} and the ``oval disk'' first noted by \citet{kormendy79}; see \citet{schinnerer00} for evidence that the latter structure is dynamically barlike.  Because the galaxy is so large in angular size, we combined a total of three adjacent SDSS fields.  The center is somewhat dusty, so that we used a $z$ image to determine the center, and then translated this information to the \emph{r} image. Although we derive a somewhat higher inclination and a different PA than did \nocite{pt06}PT06, we get the same classification (this is the only overlap of our sample with PT06).

{\bf NGC~1161 (I; S0$^0$)}: 
\citet{franco03} suggested for this object the classification SAB0 pec. However, we find no evidence for a bar in this galaxy. Their suggestion might be due to the presence of a weak dust lane at $\sim$ 8\arcsec{} (1.1 kpc) from the center of the galaxy. No break in the profile is seen out to at least 4 scalelengths.

{\bf NGC~2300 (I; SA0$^0$)}: 
Some observers have classified this as an elliptical (\citet{huchtmeier94} classified it as E3), but its profile belies this. Asymmetries are seen in the outer disk, notably in the NE where a lobe is seen at $\sim 170$\arcsec{} (some 24 kpc) from the center, possibly due to an interaction. NGC~2276 is optically close (at 7.6\arcmin), though the two redshifts ($\sim 1905$ km s$^{-1}$ and $\sim 2410$ km s$^{-1}$) are sufficiently different to cast doubt on an interaction \citep{sandage94}. In the same direction from the center as the extended lobe, but at a slightly smaller radius (90\arcsec{}), \citet{fabbiano92} found a source of X-ray emission with no optical counterpart. The profile is of Type I and it reaches $\sim$ 530\arcsec{} (75 kpc) from the center before reaching our 4.94  $\sigma_{\rm sky}$ uncertainty limit. No break in the profile is seen out to at least 4 scalelengths. 

{\bf NGC~2460 (II-CT; SA(s)a)}: 
This galaxy has a tightly wound multi-armed spiral structure starting deep inside the galaxy. Even in what should be the genuine bulge there is a lot of structure. In the outer zone the two arms join to form a single arm, yielding an unusual appearance. This seems to be associated with the surface brightness truncation.  There are no signs of a bar or a ring.  

{\bf NGC~2775 (III-d(?); SA(r)ab)}:
In this galaxy there is a bright nucleus in an elliptical genuine bulge. The disk extends in a multi-armed structure out to 85\arcsec{} (7.4 kpc) from the center, where there is a band of dust absorption. From this point outwards the galaxy continues without fine structure forming what may be a stellar halo. This idea is backed up by the isophotes, whose ellipticity declines from 85\arcsec{} (11.2 kpc) reaching a minimum at 213\arcsec{} (18.5 kpc). Nevertheless the ellipticity increases again and the surface brightness profile is an almost perfect exponential between 138\arcsec{} (12 kpc) and 345\arcsec{} (30 kpc), suggesting that it has disk structure. This is why we have classified the profile as Type III-d(?), with the question mark. The radius where the fine structure ends coincides with the point where the profile breaks, and the antitruncated disk begins. 

{\bf NGC~2985 (III-d; (R')SA(rs)ab)}:
The central part of the disk (out to 43\arcsec{}) has a tightly wound multi-armed structure, followed by another weaker spiral which extends out to 107\arcsec{} (11 kpc). From there on out the spiral arms are more tenuous, forming a pseudo-ring near 68\arcsec{} (7 kpc). The image quality does not let us trace the profile reliably further out than 312\arcsec{} (32 kpc), but the profile is clearly Type III.
Since the ellipticity at $r \sim 150$--200\arcsec{} is only marginally \textit{higher} than the ellipticity further inside, and since spiral structure is clearly visible beyond the break radius, we classify the profile as III-d. 

{\bf NGC~3031 (II.o-OLR; SA(s)ab)}: 
This galaxy is the nearest in our sample, at 3.6 Mpc. We found a bar of $\sim$ 17\arcsec{} (300 pc) radius which coincides with a maximum in the ellipticity; this is the small bar previously reported by \citet{elmegreen95}. 
Inspection and ellipse-fitting of Spitzer IRAC images \citep[Program ID 1035;][]{willner04} provides evidence for a bar with 136\arcsec{} (2.4 kpc) as the semi-major axis of maximum isophotal ellipticity (\amax{}) and 216\arcsec{} (3.8 kpc) as an upper limit on the bar size (\lbar{}), with the latter estimated from the apparent inner ring in GALEX images \citep{gdp07}.  (The boxy morphology in this region also suggests a bar; see 
Erwin \& Debattista, in prep.) Two outer arms open up quickly from a radius of 250\arcsec{} (4.4 kpc), forming an incomplete outer pseudoring. We measured the outer pseudoring on both a VLA 21 cm image \citep{adler96} and a GALEX NUV image, both available from NED.  Since the SE arm of the outer pseudoring is more symmetric, and can be seen to wrap around onto the NW arm, we measured from the nucleus to the center of the SE arm at its maximum extent.  In both images, this yielded a semi-major axis of $\sim 570\arcsec{}$ ($\sim 10$ kpc), which we adopted as the radius for the outer pseudoring.  Since the break in the surface brightness profile is only slightly inside (and a matching bump can be seen on top of the underlying broken exponential at $r \sim 600\arcsec{}$), this is a good example of an OLR break.

The outer-disk orientation for this galaxy (PA $= 150\arcdeg{}$, $i = 58\arcdeg{}$) is based on the analysis of \hi{} data in \citet{adler96}; almost identical values can be found using the large-scale ($r \gtrsim 1000\arcsec{}$) isophotes.

{\bf NGC~3032 (I; SAB(r)0$^0$)}:  
Although this galaxy is classified in the NED as SAB(r)0$^0$ there is no observable bar; see the discussion in \citet{erwin03a}. There is a weak ring between 15\arcsec{} (1.6 kpc) and 22\arcsec{} (2.3 kpc), and another peak in ellipticity close to 32\arcsec{} (3.3 kpc) which could be due to a weak arm-like feature. The nucleus is very bright and point-like. This is the only galaxy for which we combined exposures from different runs. No break in the profile is seen out to at least 5 scalelengths.

{\bf NGC~3169 (I; SA(s)a pec)}:  
This is a peculiar galaxy, clearly interacting with NGC~3166 \citep{mollenhoff01}, as seen from the weak low brightness bridge. Its bulge is large, though with considerable dust, and its spiral structure is filamentary. In spite of its irregularities, the brightness profile is of Type I. \citet{eskridge02} did mention ``evidence for a weak bar'', but we do not find any sign of this, in agreement with \citet{laurikainen04} and \citet{menendez-delmestre07}. No break in the profile is seen out to at least 6 scalelengths. 

{\bf NGC~3245 (III-s; SA(r)0$^0$)}: 
The ellipticity profile shows a peak at 15\arcsec{} (1.5 kpc), which coincides with a ring with inner and outer semiaxis lengths 12\arcsec{} and 1.9\arcsec{}, respectively. This ring appears diffusely distributed along the minor axis (\nocite{michard94}Michard \& Marchal 1994 call it a ``curious grey asymmetry'') so that it might perhaps be considered a bar, but this is not clear. The surface brightness profile shows an antitruncation with break point at 120\arcsec{} (12 kpc). The ellipticity decreases continuously, from $\sim 0.48$ at 50\arcsec{} (5 kpc) to 0.2 at 160\arcsec{} (16 kpc), so we classify this as Type III-s.


{\bf NGC~3455 (II-CT + III-d; (R')SAB(rs)b)}:  
Although this galaxy is classified as SAB, we were unable to find any convincing evidence for a bar.  The ellipticity of the isophotes stays roughly constant out to at least $\sim$ 100\arcsec{} (7.5 kpc), well beyond the antitruncation break ($R_{\rm brk}$ = 40\arcsec{}); in addition, two spiral arms extend out to $r \sim$ 70\arcsec{} (5.4 kpc). Thus, we are confident that the outer profile is III-d. 

{\bf NCG 3599 (I; SA0$^0$)}: 
This galaxy, almost face-on, has a weak ring between 45\arcsec{} (4.3 kpc) and 71\arcsec{} (6.8 kpc) and a small bar $\sim$ 11 \arcsec{} (1.05 kpc) long (see Figure~\ref{fig:n3599}). No break in the profile is seen out to at least 6 scalelengths.

{\bf NGC~3604 (III-d(?); SA(s)a pec)}:  
The inner region of this galaxy is dusty and slightly asymmetric, while one or more arcs and a spectacular off-center tidal tail dominate the outer regions, indicating that this galaxy is in the midst of an interaction. The outer profile is clearly affected by the arcs and tidal arm; we tentatively classify the profile as III-d only because the outer light is so clearly \textit{not} anything like a ``spheroid''; it is conceivable that the tidal structures could eventually evolve into something more spheroid-like. This antitruncation might possibly be caused by UGC~6306 at 3 arcminutes from NGC~3604, if the measured radial velocity difference (170 \kms{}) is due to orbital motion rather than the Hubble flow.%

{\bf NGC~3607 (I; SA(s)0$^0$)}:  
The circular dust lane does not appear to have a significant influence on the brightness profile. The central region of the bulge is structured, and may contain an internal disk. However this is not resolvable in our image. No break in the profile is seen out to at least 4 scalelengths. 

{\bf NGC~3619 (III; (R)SA(s)0$^+$)}: 
This is probably a nearly face-on shell galaxy. Four shells, which show up as arcs, can be seen, two at 129\arcsec{} (14.9 kpc) and two weaker arcs at 185\arcsec{} (21.3 kpc) and 200\arcsec{} (23.1 kpc). One of the two inner arcs might be the extension of an arm. This structure makes it difficult to determine the galaxy's orientation, though the inner isophotes are very round and the galaxy is probably close to face-on.  The surface brightness profile is clearly of Type III, with a gradual change of gradient around the break; the outer slope is the region where the shells/arcs show up. 

{\bf NGC~3626 (I; (R)SA(rs)0$^+$)}: 
Although classified as unbarred, this galaxy has a broad bar structure of radial length between 21\arcsec{} (2 kpc)  and 47\arcsec{} (4.4 kpc) (Figure~\ref{fig:n3626}), surrounded by a pseudo-ring at 71\arcsec{} (6.7 kpc) radius. It also has a secondary bar of radius $\sim$ 4\arcsec{} (400 pc), along with a ring at 5.2\arcsec{} (490 pc) and a clear dust ring at 15\arcsec{} (1.4 kpc). No break in the profile is seen out to at least 6 scalelengths.  


{\bf NGC~3675 (III-s; SA(s)b)}: 
Although \citet{eskridge00} say that this galaxy shows a strong bar in the IR, \citet{mollenhoff01} find no indication of a bar in their JHK observations, and we find no signs of a bar in either the SDSS images or any of the publically available near-IR images. A very bright nucleus is surrounded by a spiral disk, with a dust structure which shows up more strongly in the East, suggesting a possible warp. The profile is Type III with a break at $\sim$ 154\arcsec{} (9.6 kpc); since the ellipticity decreases steadily out to the limits of our free-ellipse fitting (ellipticity = 0.3 at $r \sim$ 270\arcsec{}), we classify this as III-s. %

{\bf NGC~3813 (II-CT+III-s; SA(rs)b)}: 
This is a small spiral, of irregular form with a bright nucleus. In the spiral structure there are bright knots. The brightness profile is quite noisy beyond 160\arcsec{} (17 kpc). There is a sharp downward break in the profile slope at 43\arcsec{} (4.6 kpc), making for a Type II profile.  However, beyond $\sim$ 70\arcsec{} (7.5 kpc), still used for the fit, the profile becomes shallower, accompanied by much rounder isophotes (the ellipticity drops from $> 0.6$ to $\sim 0.25$ between 50\arcsec{} and 65\arcsec{}), so the outer profile is pretty clearly III-s. %

{\bf NGC~3898 (III-d; SA(s)ab)}: 
A bright nucleus sits within a bulge well defined out to 27\arcsec{} (2.5 kpc) on the SDSS image. From this point outwards we see a clear disk, with a multi-armed spiral structure out to some 90\arcsec{} (8 kpc). The surface brightness is very low beyond this radius, but there are clear signs of an arm going out to beyond 220\arcsec{} (20 kpc). This outer zone retains the ellipticity of the inner ($\sim 0.4$). So we consider it an extended part of the disk; thus, this is a Type III-d profile. The break is found at $r \sim$ 111\arcsec{}.

{\bf NGC~3900 (III-d; SA(r)0$^+$)}:   
A ring between 31\arcsec{} (3.9 kpc) and 47\arcsec{} (5.9 kpc) bounds the denser part of the disk, surrounded by a fainter zone with a spiral form. The ring shows up as a feature with local changes in gradient on the brightness profile.  The isophotes remain highly elliptical beyond the break; for $r >$ 240\arcsec{} (30 kpc), they actually become \textit{more} elliptical, which might hint at a warp or interaction affecting the outer disk. The broad, smooth transition zone between the inner and outer exponentials is similar to that seen for the Type~III-d profiles of NGC~3489 and NGC~7177 in Paper~I. It could indicate that we are seeing two co-extensive, superimposed components, though the high ellipticity of the isophotes implies that both components have a similar flattening. 

{\bf NGC~3998 (III-d(?); SA(r)0$^0$?)}: 
In this object we see a small bar ($\sim$ 7.8\arcsec{}; 0.6
kpc) and a ring between 30\arcsec{} (2.1 kpc) and 50\arcsec{} (3.5 kpc) 
which produces a slight bump on
the brightness profile, and coincides with a peak in the ellipticity at
36\arcsec{} (2.4 kpc). \citet{sanchez04} claimed that this galaxy has a Freeman Type
II disk, with a cut-off at 30\arcsec{} (2 kpc), but this is probably just
the effect of the bump associated with the ring.  Our profile extends
out to $\sim$ 350\arcsec{} (23 kpc) and shows an antitruncation with its break point at
120\arcsec{} (8 kpc). As this galaxy is only slightly inclined, so that its ellipticity
is low over the whole of the disk, both inner and outer, it is not easy
to say that there is no surrounding stellar halo. However, the fact that
isophotes outside the break show no significant decline in ellipticity
leads us to classify this as Type III-d(?), albeit with some uncertainty.

{\bf NGC~4138 (III-d; SA(r)0$^+$)}:  
The bulge of this galaxy is surrounded by a dust lane, which subtends almost a semi-circle at the center. The contrast between this dust feature and the bright spiral structure arising from the bulge can explain the ``ring'' classification of this object. In the brightest zone, at $\sim$ 20\arcsec{} (1.3 kpc) radius the ellipticity peaks, and there is slight bump in the brightness profile. The profile as a whole is of Type III, with a sharp change of slope at the break point. Traces of spiral structure can be seen out to approximately the break radius; the ellipticity of the isophotes remains almost perfectly constant out to at least 140\arcsec{} (9 kpc), making this a clear III-d profile. We agree with \citet{sandage94} that S0  is not the best classification for this galaxy. However, we want to retain the original classification here for purposes of discussing RC3 Hubble type statistics. %

{\bf NGC~4150 (III-s; SA(r)0$^0$)}: 
The innermost isophotes of this galaxy are distorted by a roughly circular dust lane (radius $\sim 6.6\arcsec$); there is fine spiral and irregular dust structure at smaller radii, clearly visible in HST images \citep[e.g.,][]{lauer95,quillen00}.
The ellipticity declines steadily for $r \gtrsim $ 45\arcsec{} (3 kpc), which suggests gradual domination of the light by a rounder outer structure, hence the III-s profile classification. %

{\bf NGC~4223 (III-s(?); SA(s)0$^+$)}:  
The bulge is encircled by a very tight spiral forming a pseudo-ring between 35\arcsec{} and 46\arcsec{} (2.8 and 3.7 kpc) in radius. Further out, at $\sim $ 80\arcsec{} (6.5 kpc), we see two extended arms in an elongated \emph{S} shape, which cause the isophotes to become more elliptical.  The isophotes appear to become rounder for $r > $ 100\arcsec{} (8 kpc) (though we cannot measure isophote shapes at $r \gtrsim $ 190\arcsec{}), which leads us to tentatively classify this profile as III-s.  

{\bf NGC~4281 (III-s; S0$^+$)}: 
The ellipticity of this galaxy rises to a relatively constant plateau value (the latter covering the range 22\arcsec{} -- 55\arcsec{}), 
and then falls smoothly at larger radii.  This suggests an inclined disk embedded in a rounder, luminous spheroid, making this a clear if somewhat extreme example of a Type III-s profile.  There is a small dust ring with radius $\sim$ 9\arcsec{} (1 kpc), which produces a dip in the ellipticity profile. 

{\bf NGC~4369 (II.o + III; (R)SA(rs)a)}: 
Although classified as unbarred, near-IR images show a very strong, short bar \citep{knapen00}.  In our image we detect a ring between 36\arcsec{} and 57\arcsec{}. 
The coincidence of this ring with the Type II break might suggest an OLR break; however, the break is at $\sim 5$ times the bar radius, which is well outside the usual range seen in \nocite{epb08}Paper~I. Consequently, we leave the inner classification at II.o. Since the galaxy is essentially face-on, with no visible evidence for spirals beyond the break at $r \sim $ 112\arcsec{} (9 kpc), we are unable to sub-classify the outer part of the profile.  


{\bf NGC~4459 (III-d; SA(r)0$^+$)}: 
This object has a bright nucleus and a dust ring out to 8\arcsec{} (620 pc). In fact this ``ring'' has a very fine multi-armed spiral structure, 
clearly visible in HST ACS images \citep[see Fig.~7 of][]{ferrarese06}.
This structure gives the outer part of the bulge (which should really be termed a pseudobulge; \nocite{kormendy04}Kormendy \& Kennicutt\ 2004) a ring-like appearance and a minimum in the ellipticity, as well as a small jump in the brightness profile. From 55\arcsec{} (4.3 kpc) to 270\arcsec{} (21 kpc) the profile is a double exponential, with the break at 120\arcsec{} (9.3 kpc); since the ellipticity is roughly constant out to at least 200\arcsec{} (16 kpc), we judge this to be a III-d profile. \citet{michard94} refer to this galaxy as a ``well-known S0 dominated by a disk''.  

{\bf NGC~4578 (I; SA(r)0$^0$)}: 
A broad, weak ring between 50\arcsec{} (4 kpc) and 90\arcsec{} (7 kpc) produces a clear bump in the surface brightness profile; the underlying profile is clearly Type I. 
No break in the profile is seen out to at least 6 scalelengths.


{\bf NGC~4736 (II.o-OLR; (R)SA(r)ab)}: 
This is one of the two cases where we have made a mosaic of INT-WFC images, since this galaxy is very large. Although traditionally classified as unbarred, NGC~4736 has in fact two bars of  semi-major axis lengths 170\arcsec{} (4.2 kpc) and 25\arcsec{} (0.6 kpc), respectively.  The outer bar is weak, and has been classified by some authors as an ``oval disk''. However, there is evidence that its dynamical behavior is that of a bar \citep[see, e.g., the discussion in][]{erwin04}. This is a similar system to NGC~1068, although the outer pseudoring in NGC~4736 is relatively faint. 

{\bf NGC~4750 (III-d; (R)SA(rs)ab)}: 
There is a ring formed by tightly wound spiral
structure at some 16\arcsec{} (2 kpc) radius and a similar structure at $\sim$ 45\arcsec{} (5.5 kpc)
radius. There are indications in the optical images of a weak bar of
length some $\sim$ 4\arcsec{} (490 pc); this is consistent with the known nuclear
bar detected in the near-IR by \citet{laine02}. These features produce
some waviness in the inner part of the brightness profile.  See
Section~\ref{sec:bars} for discussion of the weak, large-scale bar. In
the outer zone the disk is elongated N-S, and at the extremities  we
detect material apparently being accreted onto the disk; the
antitruncated profile may be due to this accreting material. 

\textbf{NGC~4772 (I; SA(s)a)}: 
Evidence for a very large, faint bar in the galaxy is discussed in Section~\ref{sec:bars}.  
Our best estimate for the galaxy's overall orientation using the outermost low S/N isophotes in the SDSS image is consistent with that from \citet{haynes00}.
Given that the bar extends to at least $\sim$ 85\arcsec{} (6 kpc) in radius, we consider the profile out to $\sim$ 110\arcsec{} (8 kpc) to be part of the extended excess light associated with the bar (as in other barred galaxies profiles, e.g., NGC~278, 4369, and 4750, or numerous examples in \nocite{epb08}Paper~I).  Consequently, we fit the profile outside that radius as the outer disk proper; the result is a Type I profile.  The broad bump in the profile centered at $r \sim$ 200\arcsec{} (14 kpc) appears to correspond with the outer \hi{} ring reported by \citet{haynes00}; see the discussion in Section~\ref{sec:bars} and Figure~\ref{fig:n4772}. 
No break in the profile is seen out to at least 5 scalelengths.


{\bf NGC~4826 (III-s; (R)SA(rs)ab)}: 
This is the second of two galaxies for which we had to create a complete mosaic from individual fields of the INT-WFC. It has an outer ring of $\sim $ 180\arcsec{} (6.3 kpc) radius. The strong dust structure within this object causes the brightness profile to be somewhat wavy at $\sim$ 60\arcsec{} (2 kpc).  The ellipticity of the isophotes drops sharply beyond $r \sim$ 340\arcsec{} (12 kpc), coinciding with the appearance of the outer, shallower profile, which is good evidence for a rounder spheroid surrounding the disk proper.

{\bf NGC~4880 (II-CT; SA(r)0$^+$)}: 
The surface brightness profile shows a truncation at $\sim$ 68\arcsec{} (6.5 kpc). The inner disk has spiral structure with thick tightly wound but not very well defined arms. Given the dust in the inner disk, it is difficult to completely rule out a weak bar in this galaxy, but in the absence of stronger evidence we consider this galaxy to be unbarred. 

{\bf NGC~4941 (I; (R)SAB(r)ab)}: 
On this image we can see, between 37\arcsec{} and 73\arcsec{}, 
a bright annular region which is clearly a disk with strong streaks of dust in front of it. We can also see a broad but weak bar (out to 95\arcsec{}; see Figure~\ref{fig:n4941})
at the extremities of which there is an outer ring, between 81\arcsec{} (5.9 kpc) and 120\arcsec{} (8.7 kpc) \citep[this bar plus ring structure was termed an ``oval disk'' by][]{kormendy79}. The ring produces a bump in surface brightness which is seen above the exponential disk component. Classification of this profile is not straightforward, but overlooking the internal waviness and concentrating on the outer part of the disk we find that we are dealing with a Type I profile. No break in the profile is seen out to at least 6 scalelengths.

{\bf NGC~5273 (II-CT + III-d; SA(s)0$^0$)}: 
Between $r \sim$ 26\arcsec{} (2 kpc) and 44\arcsec{} (3.4 kpc) there is a ring onto which two inner spiral arms converge. The form of the brightness profile makes this a Type II+III-d, but there is some ambiguity. For example, it is possible that the combination of bulge and ring produces a broad excess in the inner part of the galaxy, superposed on a Type I profile. Although the ellipticity is low, it remains almost constant out to $\sim$ 150\arcsec{} (11.7 kpc), leading us to classify the outer profile as III-d. 

{\bf NGC~5485 (I; SA0$^0$ pec)}: 
This galaxy is classified as peculiar. It has a 40-arcsec-long dust lane which crosses the bulge close to the nucleus, perpendicular to the major axis. Between 112\arcsec{} (14 kpc) and 178\arcsec{} (22 kpc) we see a very weak ring, which produces a slight, extended bump in the surface-brightness profile. Although the profile clearly extends beyond 400\arcsec{} (50 kpc), we have not measured it further out as stray light affects the outer region. No break in the profile is seen out to at least 4 scalelengths.

{\bf NGC~5520 (I; Sb)}:
This object has a dusty and asymmetric spiral structure extending out to the edge of the measured disk, but in spite of this the profile is clearly Type I. We think the bump at $\sim$ 12\arcsec{} is related to the dusty and asymmetric structure. 
No break in the profile is seen out to at least 6 scalelengths.

{\bf NGC~6340 (III; SA(s)0/a)}: This face-on galaxy shows a faint dust lane near
the nucleus coming in from the west. It has several fine tightly wound
arms wrapped around the bulge in almost circular structures, and there are a
number of dust lanes between the bulge and the rest of the disk; this 
extends outwards to large radii following an antitruncation. Traces of
stellar spiral structure can be seen in unsharp masks out to $r \sim$ 85\arcsec{}
(8.3 kpc).  Since the galaxy is approximately face-on and we do not see
any spiral structure at larger radii, we cannot determine whether the
profile outside the break is still part of the disk or not; thus, the
profile is an undefined Type III.  

{\bf NGC~7217 (III; (R)SA(r)ab)}:  The bulge of this galaxy, as normally defined (out to some 52\arcsec{}) 
is in fact filled with dust structures in loosely spiral forms. At some 72\arcsec{} (5.2 kpc) 
out from the center of this galaxy, a tightly wound spiral of filamentary arms forms a pseudo-ring 
which quickly (at $\sim$ 100\arcsec{}) 
dissolves into a smooth structure further outside. This regions shows up in the surface-brightness profile as a bump on top of the antitruncation transition.  This galaxy is too close to face-on, and too featureless in the outer regions, for us to determine whether the outer light is from a spheroid or a continuation of the disk. 

{\bf NGC~7457 (III-d; SA(rs)0$^-$?)}: There is no notable structure in this S0 galaxy, and the brightness profile is quite smooth.  It shows an exponential slope beginning at $r \sim$ 20\arcsec{} (1.3 kpc), with a break to shallower gradient at $R_{\rm brk} \sim$ 42\arcsec{} (2.6 kpc). Outside the (rounder) photometric bulge region, the ellipticity of the isophotes is virtually constant with radius, indicating that we are seeing a disk at all radii.

{\bf UGC~3580 (I; SA(s)a pec)}: In this galaxy we can see considerable structure even within the inner part of the bulge, close to the nucleus. A weak ring of ``knots'' encircles the bulge at $\sim$ 28\arcsec{} (2.6 kpc) radius. No break in the profile is seen out to at least 6 scalelengths.

{\bf UGC~4599 (III-d; (R)SA0$^0$)}: This is a face-on galaxy with striking ring composed of tightly wound spiral arms.  Spiral arms begin at a radius of $\sim$ 40\arcsec{} (6 kpc)
and wind out to form a ring with a radius of $\sim$ 55\arcsec{} (8 kpc), %
forming a clear bump in the surface brightness profile.  Faint spiral structure can be seen as far out as $\sim$ 200\arcsec{} (27 kpc) %
to the southeast; this leads us to classify the outer-disk profile as III-d.  Interior to the aforementioned ring, the galaxy is smooth, featureless, and close to circular. Given the complete absence of any barlike structure, it is tempting to speculate that this ring might be a collisionally generated structure, rather than an OLR resonance ring.  Because the available $R$-band images were of poor quality, we used a $B$-band image to trace the outer disk profile.










\clearpage

\begin{figure*}
  \centering
  \includegraphics[height=8.0in]{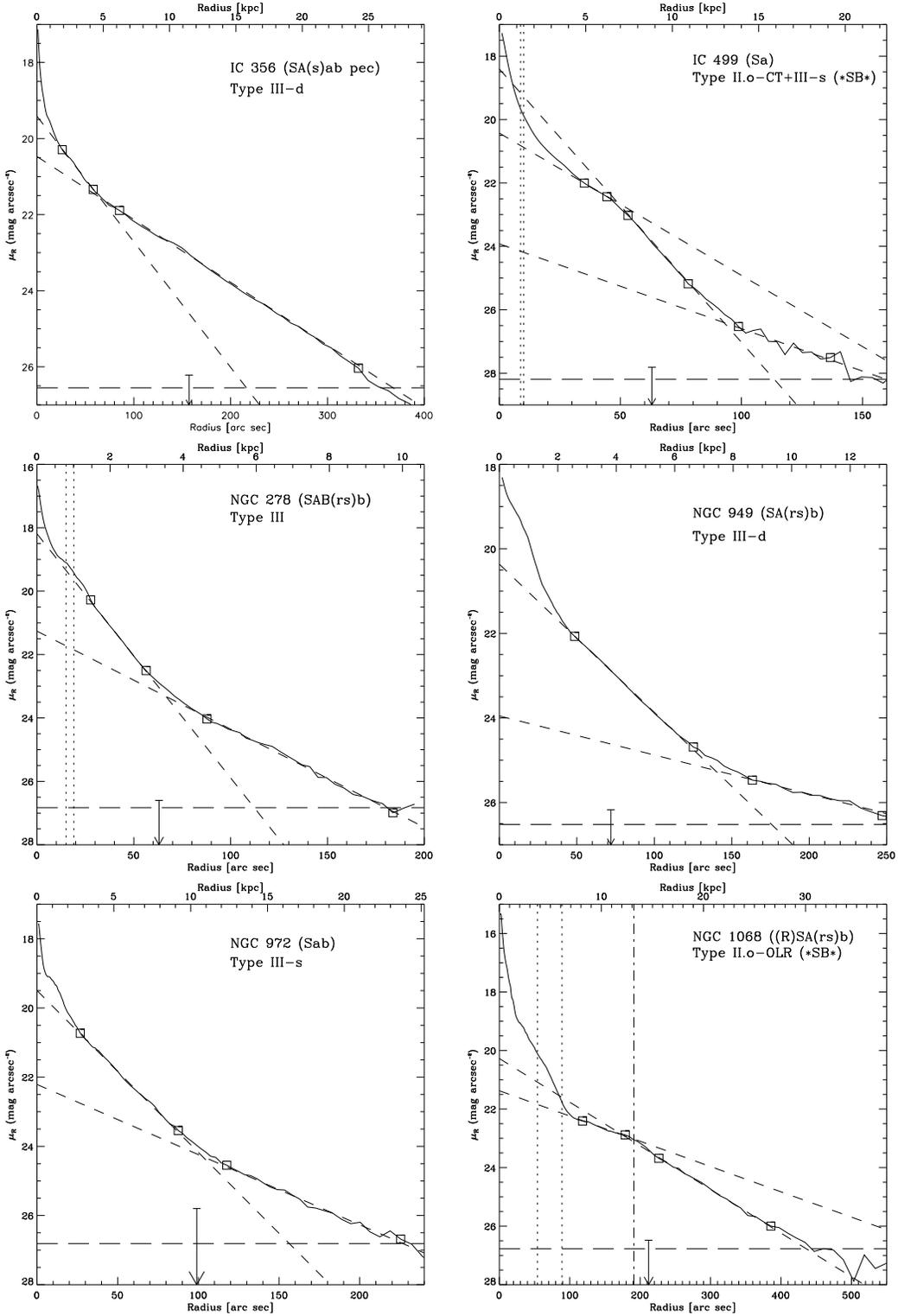}
  \caption{Azimuthally averaged radial surface brightness profiles in the $R$-band (except for UGC 4599, which is $B$-band) for the galaxies discussed in this paper.  The galaxy name, Hubble type, and disk profile type are given in the upper right of each panel. Vertical arrows indicate $R_{25}$, and dashed lines show exponential fits (a single fit for Type I, two for Type II and Type III profiles, and three fits for Type II + III), where the squares show the limits used for the fit (see Section \ref{sec:ellipse_fitting}); the sky uncertainty limit (see Section \ref{sec:sky_subtraction}) is indicated by the horizontal dashed lines.  Vertical dot-dashed lines indicate radii of prominent rings. For galaxies with bars, vertical dotted lines indicate lower and upper limits on the bar radius. 
  Only outer bars are shown for double barred galaxies.} \label{fig:profiles}
\end{figure*}  

\addtocounter{figure}{-1}
\begin{figure*}
  \centering
  \includegraphics[height=7.5in]{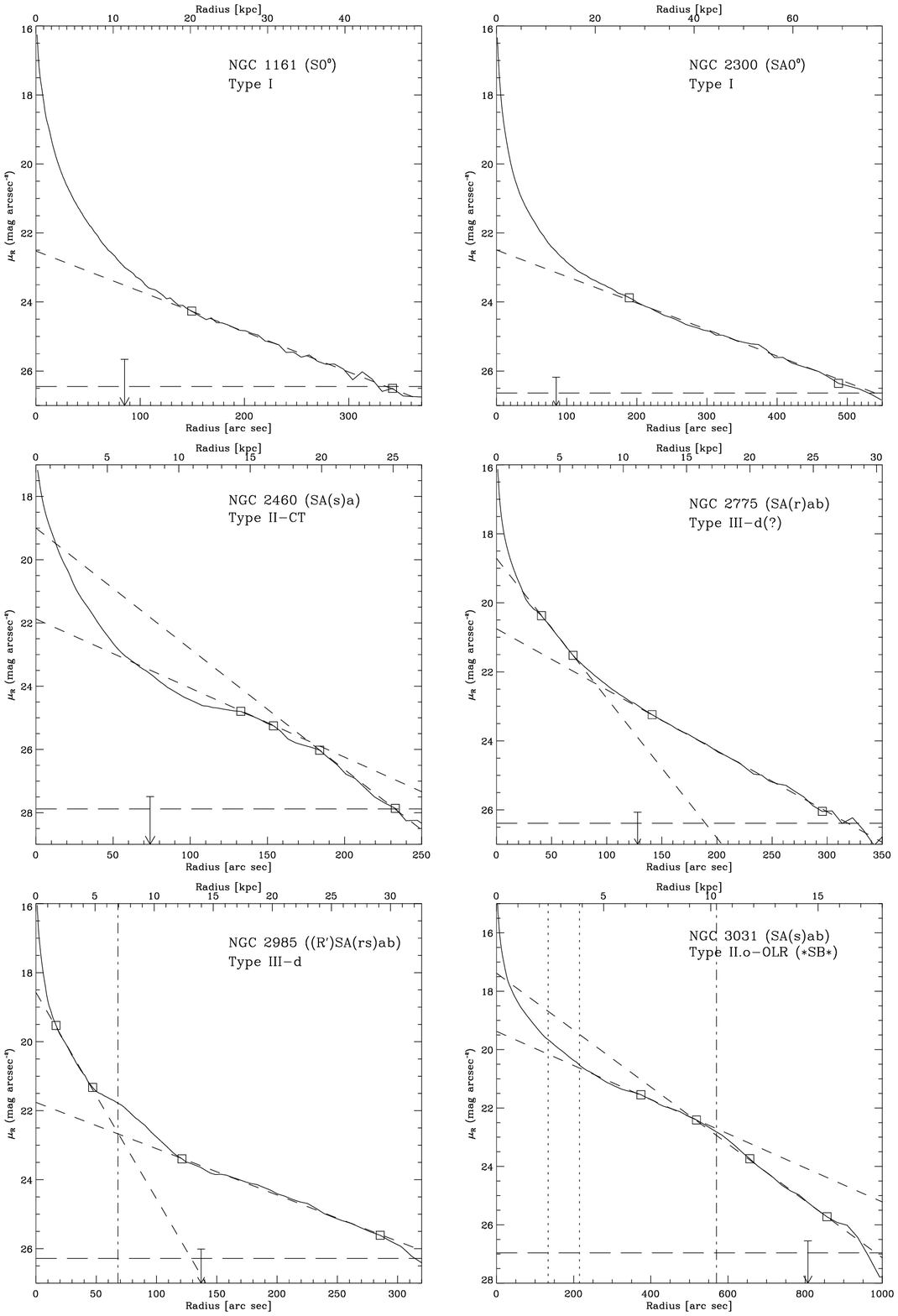}
  \caption{continued.} \label{fig:profiles_2}
\end{figure*}

\addtocounter{figure}{-1}
\begin{figure*}
  \centering
  \includegraphics[height=7.5in]{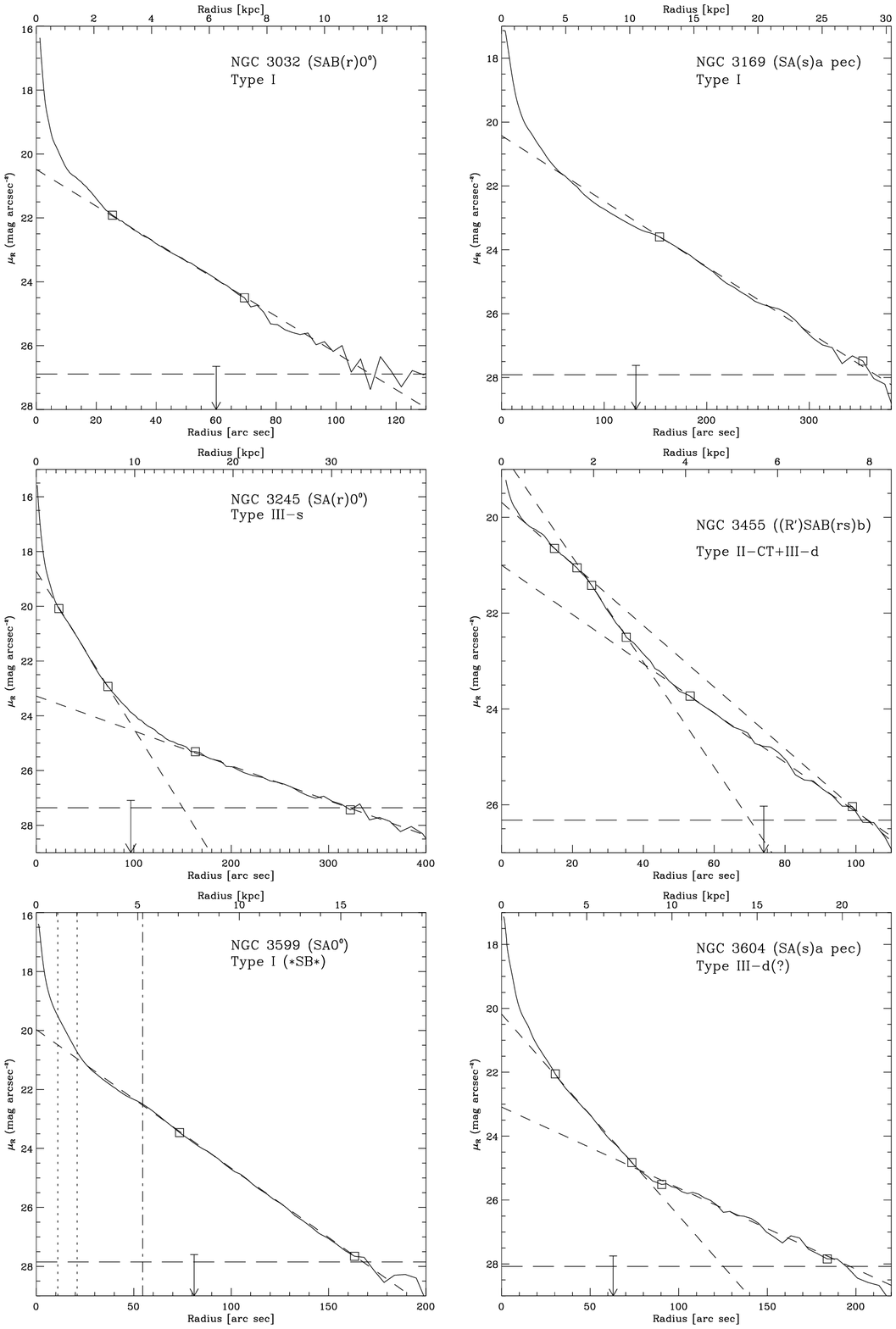}
  \caption{continued.} \label{fig:profiles_3}
\end{figure*}

\addtocounter{figure}{-1}
\begin{figure*}
  \centering
  \includegraphics[height=7.5in]{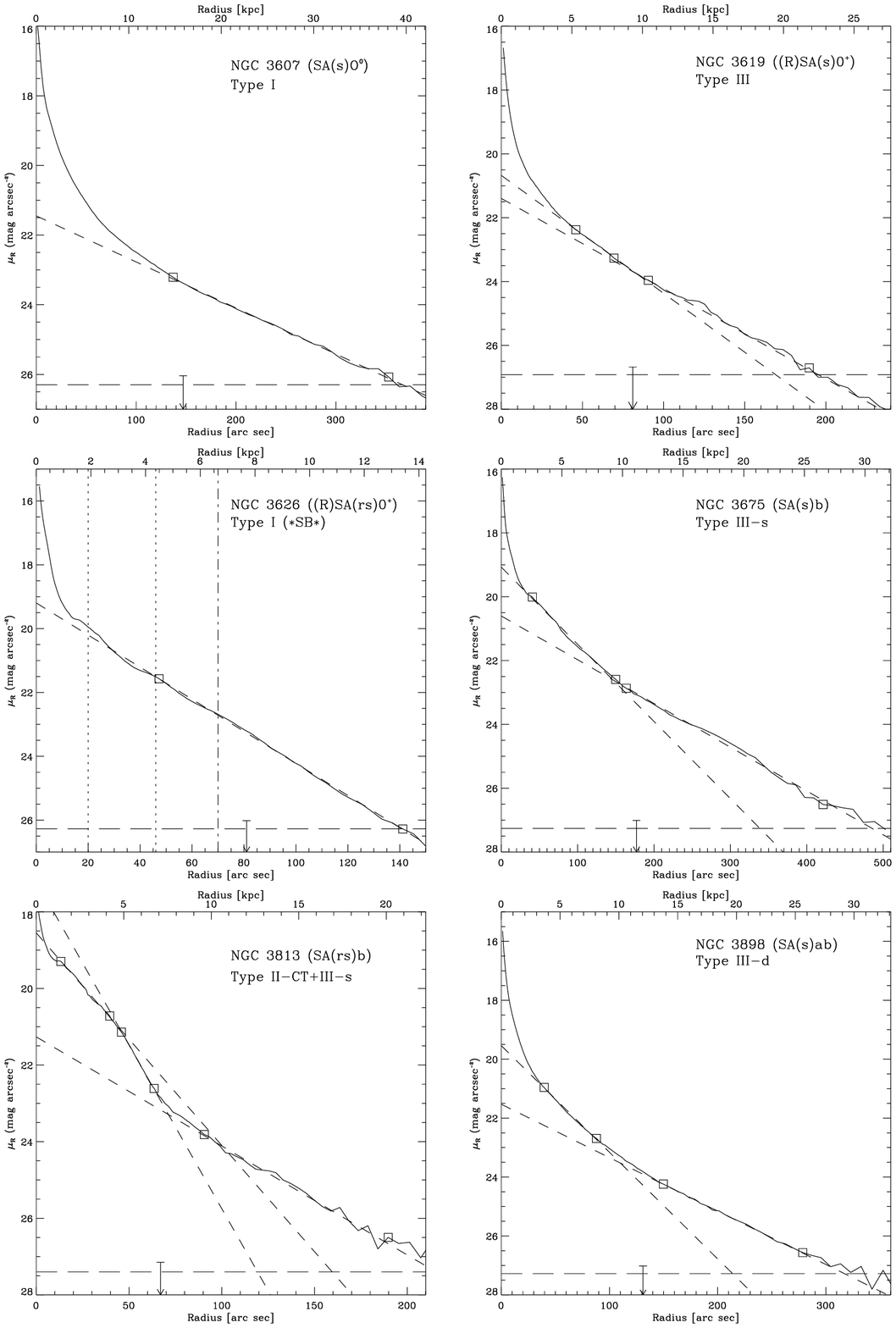}
  \caption{continued.} \label{fig:profiles_4}
\end{figure*}

\addtocounter{figure}{-1}
\begin{figure*}
  \centering
  \includegraphics[height=7.5in]{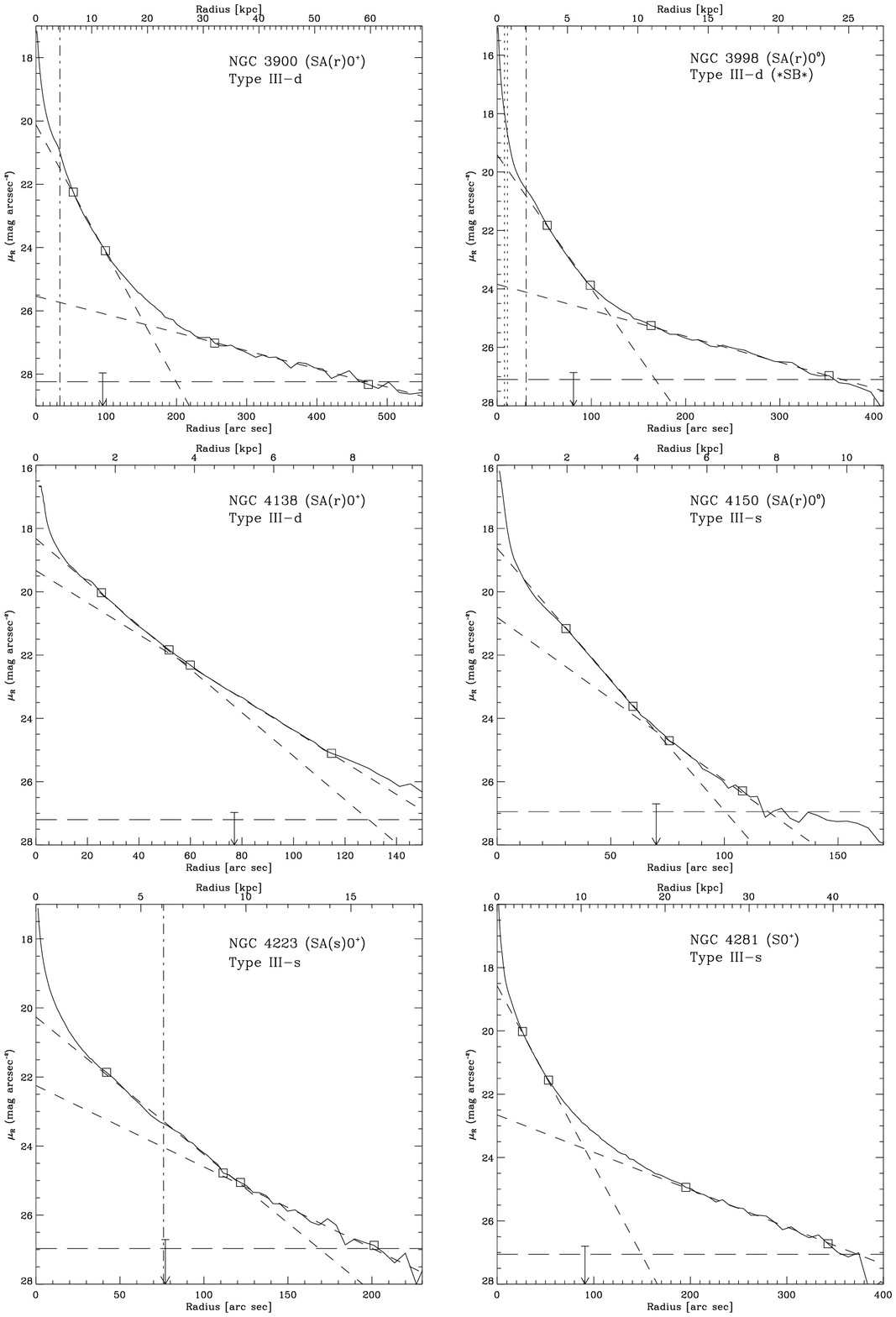}
  \caption{continued.} \label{fig:profiles_5}
\end{figure*}

\addtocounter{figure}{-1}
\begin{figure*}
  \centering
  \includegraphics[height=7.5in]{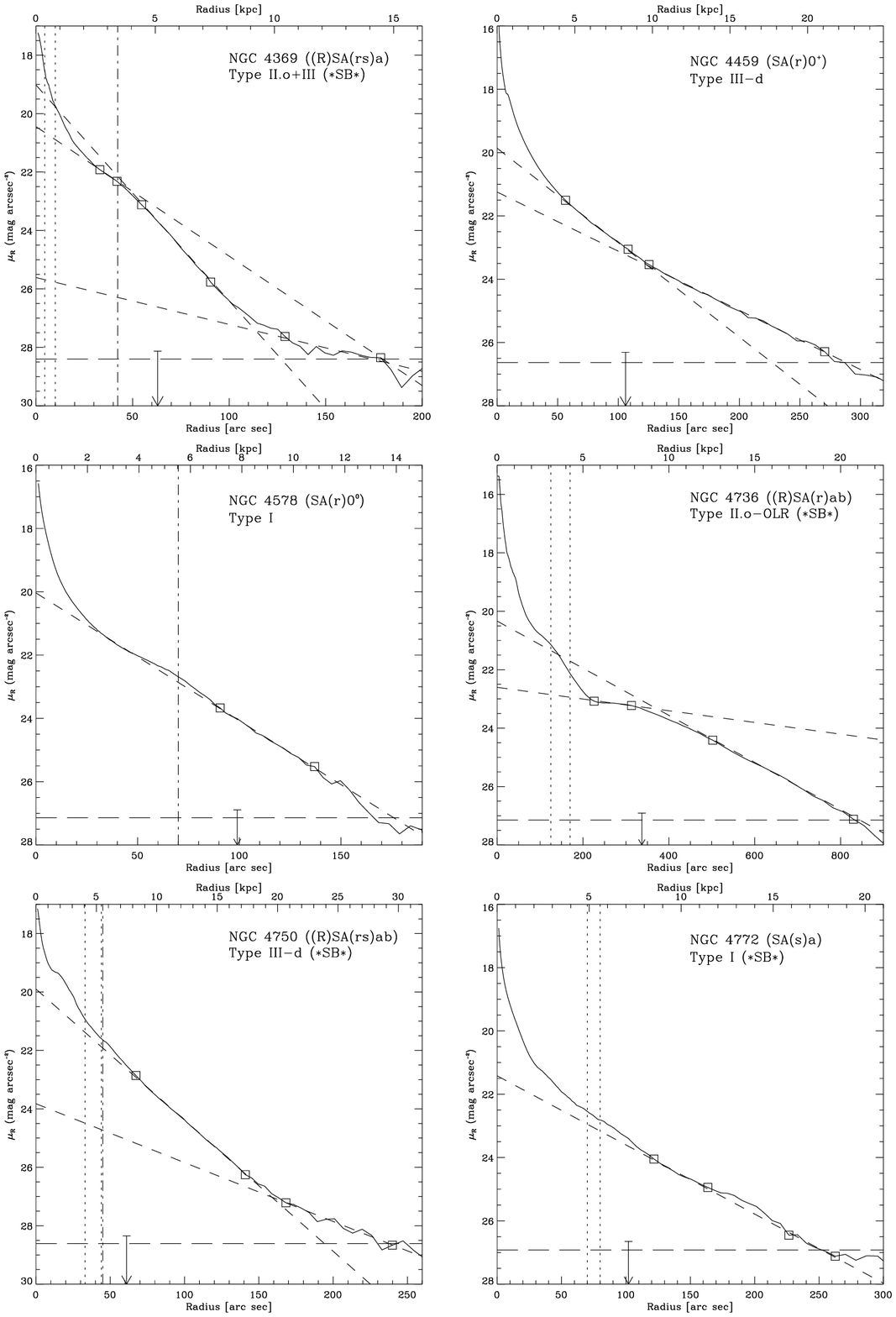}
  \caption{continued.} \label{fig:profiles_6}
\end{figure*}

\addtocounter{figure}{-1}
\begin{figure*}
  \centering
  \includegraphics[height=7.5in]{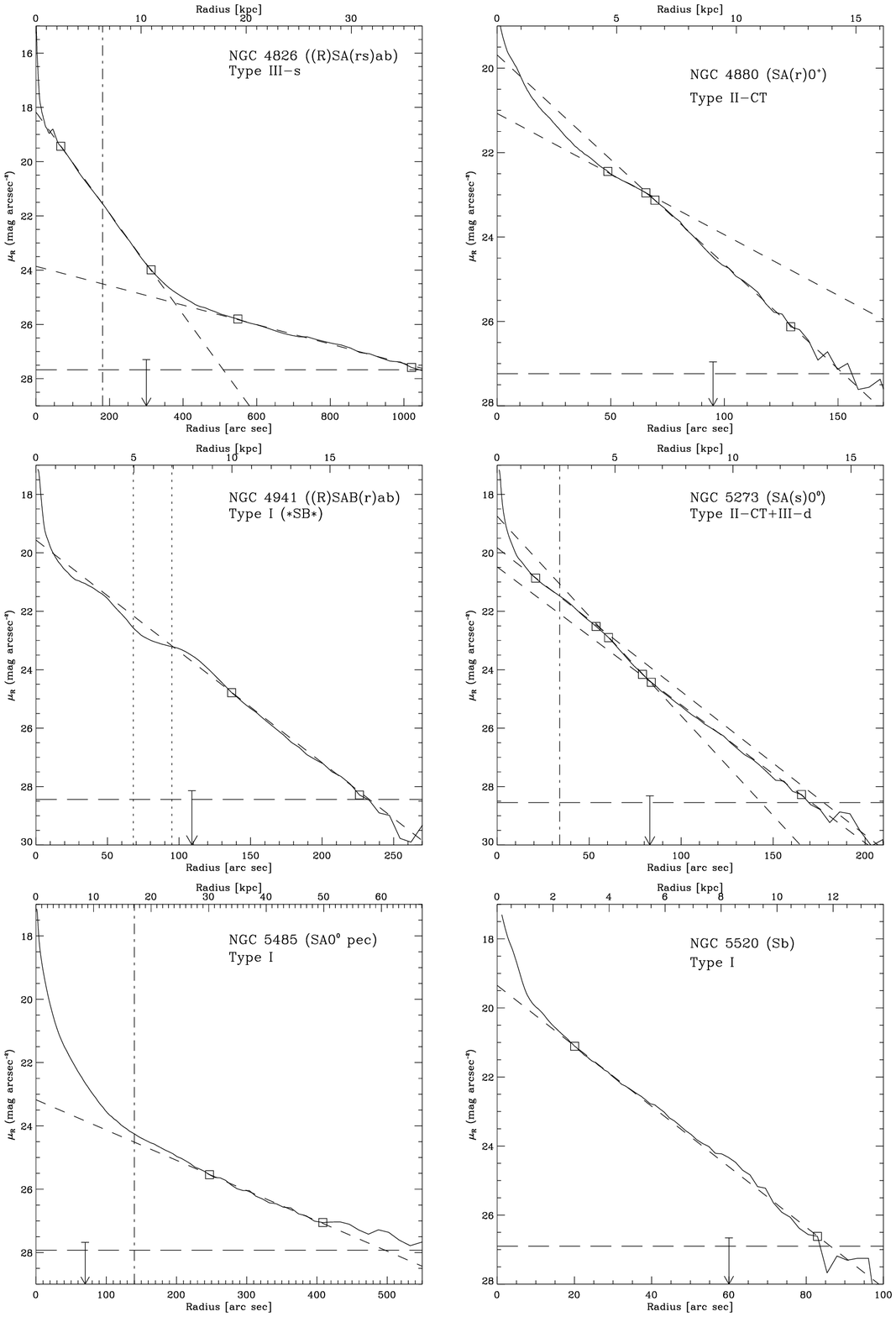}
  \caption{continued.} \label{fig:profiles_7}
\end{figure*}

\addtocounter{figure}{-1}
\begin{figure*}
  \centering
  \includegraphics[height=7.5in]{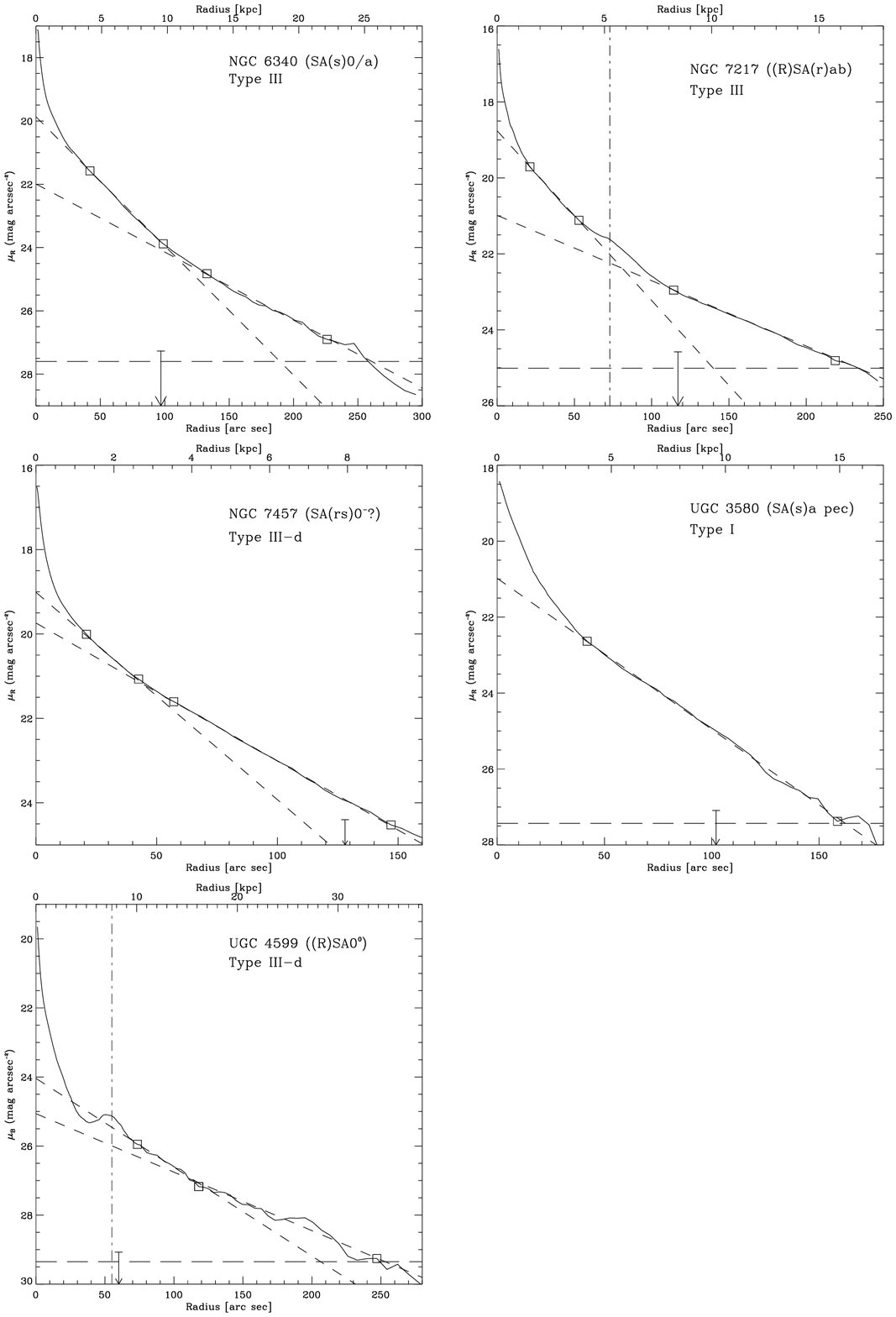}
  \caption{continued.} \label{fig:profiles_8}
\end{figure*}

\clearpage

\section{Tabulated surface brightness profiles}\label{sec:extra}
\emph{(The full tables of the brightness profiles for the whole set of galaxies listed in Tables 1, 2, 5, and 6 are available in the on-line version.)}
\renewcommand{\thetable}{B-\arabic{table}}

\setcounter{table}{1}

\begin{table}[h]
\scriptsize
\begin{minipage}[t]{0.3\linewidth}\centering
\begin{tabular}{ c c }
\multicolumn{2}{c}{IC 356}  \\
\hline  \hline 
a      &   $\mu$            \\  
arcsec & mag arcsec$^{-2}$  \\  
\hline
   1.01 & 17.13      \\
   1.04 & 17.14      \\
   1.07 & 17.15      \\
   1.10 & 17.17      \\
   1.13 & 17.18      \\
   1.17 & 17.20      \\
   1.20 & 17.22      \\
   1.24 & 17.24      \\
   1.28 & 17.26      \\
   1.32 & 17.28      \\
\nodata & \nodata  \\
\nodata & \nodata  \\
\nodata & \nodata  \\
\hline
\end{tabular}
\end{minipage}
\end{table}

\end{document}